\documentclass[%
 aip,
 amsmath,amssymb,
 reprint,%
]{revtex4-1}

\usepackage{graphicx}% Include figure files
\usepackage{dcolumn}% Align table columns on decimal point
\usepackage{bm}% bold math
\usepackage[utf8]{inputenc}
\usepackage[T1]{fontenc}
\usepackage{mathptmx}
\usepackage{etoolbox}
\usepackage{upgreek}
\usepackage{tikz}   % to be able to make the lines at the bottom
\usepackage{xcolor}
\usepackage{hyperref}

\DeclareRobustCommand{\solblack}{\raisebox{2pt}{\tikz{\draw[black,solid,line width=0.5pt](0,0) -- (3mm,0);}}} 

\DeclareRobustCommand{\solred}{\raisebox{2pt}{\tikz{\draw[red,solid,line width=0.5pt](0,0) -- (3mm,0);}}} 

\DeclareRobustCommand{\solblueone}{\raisebox{2pt}{\tikz{\draw[blue,solid,line width=0.5pt](0,0) -- (3mm,0);}}}

\definecolor{blue_colom}{rgb}{0.12157, 0.4667, 0.706}
\DeclareRobustCommand{\solblue}{\raisebox{2pt}{\tikz{\draw[blue_colom,solid,line width=0.7pt](0,0) -- (3mm,0);}}} 

\definecolor{green_colom}{rgb}{0.2392, 0.6549, 0.2392}
\DeclareRobustCommand{\dotgreen}{\raisebox{2pt}{\tikz{\draw[green_colom,dotted,line width=1pt](0,0) -- (4mm,0);}}} 

\definecolor{orange_colom}{rgb}{1, 0.4980, 0.0549}
\DeclareRobustCommand{\dashorange}{\raisebox{2pt}{\tikz{\draw[orange_colom,dashed,line width=0.75pt](0,0) -- (5mm,0);}}} 

\definecolor{red_colom}{rgb}{1, 0.0, 0.0}
\DeclareRobustCommand{\dashred}{\raisebox{2pt}{\tikz{\draw[red_colom,dashed,line width=0.75pt](0,0) -- (5mm,0);}}} 

\DeclareRobustCommand{\solorange}{\raisebox{2pt}{\tikz{\draw[orange_colom,solid,line width=0.5pt](0,0) -- (3mm,0);}}}

\DeclareRobustCommand{\grayarrow}{%
  \raisebox{2pt}{%
    \tikz{\draw[gray,solid,line width=1pt,->,>=latex] (0,0) -- (5mm,0);}%
  }%
}

\DeclareRobustCommand{\graydashedarrow}{%
  \raisebox{2pt}{%
    \tikz{\draw[gray,dashed,line width=1pt,->,>=latex] (0,0) -- (5mm,0);}%
  }%
}

\makeatletter
\def\@email#1#2{%
 \endgroup
 \patchcmd{\titleblock@produce}
  {\frontmatter@RRAPformat}
  {\frontmatter@RRAPformat{\produce@RRAP{*#1\href{mailto:#2}{#2}}}\frontmatter@RRAPformat}
  {}{}
}%
\makeatother
\begin{document}

\preprint{AIP/123-QED}

%%%%%%%%%%%%%%%%%%%%%%%%%%%%%%%%%%%%%%%%%%%%%%%%%%%%%%%%%%%%%%%%%%%%%%%%%%%%%%%%%%%%
\title{Design and Implementation of a Fast-Sweeping Langmuir Probe Diagnostic for DC Arc Jet Environments}
%%%%%%%%%%%%%%%%%%%%%%%%%%%%%%%%%%%%%%%%%%%%%%%%%%%%%%%%%%%%%%%%%%%%%%%%%%%%%%%%%%%%

% Force line breaks with \\
\author{Sebastian V. Colom}
 \email{sebastian.v.colom@nasa.gov, magnus.haw@nasa.gov, jocelino.rodrigues@nasa.gov}
 \affiliation{ 
AMA Inc. at NASA Ames Research Center, Moffett Field, CA 90435%\\This line break forced with \textbackslash\textbackslash
}%

\author{Magnus A. Haw}%
\affiliation{ 
NASA Ames Research Center, Moffett Field, CA 94035%\\This line break forced with \textbackslash\textbackslash
}%

\author{Jocelino Rodrigues}%
\affiliation{ 
NASA Postdoctoral Program at NASA Ames Research Center, Moffett Field, CA 94035%\\This line break forced with \textbackslash\textbackslash
}%

%%%%%%%%%%%%%%%%%%%%%%%%%%%%%%%%%%%%%%%%%%%%%%%%%%%%%%%%%%%%%%%%%%%%%%%%%%%%%%%%%%%%

\date{\today}% It is always \today, today,

%%%%%%%%%%%%%%%%%%%%%%%%%%%%%%%%%%%%%%%%%%%%%%%%%%%%%%%%%%%%%%%%%%%%%%%%%%%%%%%%%%%%
\begin{abstract}
Langmuir probe diagnostics are a cornerstone of plasma characterization, providing critical measurements of electron temperature, electron density, and plasma potential. However, conventional swept Langmuir probes and other traditional electrostatic probes often lack the temporal resolution necessary to capture transient plasma behavior in dynamic environments. This paper presents the design and implementation of a fast-sweeping Langmuir probe system that is open-source, low-cost, and adaptable for a wide range of plasma applications. The probe system incorporates voltage sweeping to resolve rapid fluctuations in plasma parameters at a temporal resolution of up to 200~kHz. To validate its performance, the system was implemented in the 30~kW miniature Arc jet Research Chamber (mARC~II), a high-enthalpy DC arc jet facility designed for prototype testing and development. Experimental results demonstrate the probe’s capability to operate in extreme aerothermal conditions, providing time-resolved electron temperature and density along the flow's radial profile. This work establishes a robust and accessible Langmuir diagnostic solution for researchers studying transient plasma behavior in high-enthalpy environments.
\end{abstract}
%%%%%%%%%%%%%%%%%%%%%%%%%%%%%%%%%%%%%%%%%%%%%%%%%%%%%%%%%%%%%%%%%%%%%%%%%%%%%%%%%%%%

\maketitle

%%%%%%%%%%%%%%%%%%%%%%%%%%%%%%%%%%%%%%%%%%%%%%%%%%%%%%%%%%%%%%%%%%%%%%%%%%%%%%%%%%%%
\section{Introduction}
\label{sec:intro}
%%%%%%%%%%%%%%%%%%%%%%%%%%%%%%%%%%%%%%%%%%%%%%%%%%%%%%%%%%%%%%%%%%%%%%%%%%%%%%%%%%%%
\vspace{-10pt}

Langmuir diagnostics, recently crossed their 100-year anniversary since first implemented, are a method for measuring fundamental plasma parameters. These electrostatic probes provide a direct means of characterizing electron density, temperature, and plasma potential\cite{huddlestone1965langmuir, chen2016introduction}. Over the decades, Langmuir probe techniques have evolved from simple single probes to more advanced configurations such as double probes, triple probes\cite{mARC-langmuir}, and emissive probes\cite{emmissive}, enhancing measurement accuracy and adaptability in various plasma environments. The fundamental principle involves inserting a conductive electrode into the plasma and applying a voltage to measure the resulting current, which follows a characteristic response owing to electron and ion collection dynamics. By analyzing the current-voltage response, key plasma properties can be determined using theoretical models based on electron energy distributions. Today, Langmuir probes remain a cornerstone of plasma diagnostics, widely used in laboratory~\cite{hall-thruster, haw2019laboratory, efthimion2005development}, fusion\cite{fusion, katz2008experiments}, space\cite{spacecraft}, and industrial\cite{semiconductor} plasma applications due to their versatility and direct measurement capabilities. \par 

An example of a Langmuir probe is a triple probe which uses three electrodes (typically at a fixed bias voltage), offering simplicity and robustness in fluctuating plasmas. However, triple probes are limited in diagnostic range and are forced to assume Maxwellian-distribution properties~\cite{nasa-triple}. Emissive probes, used primarily for measuring plasma potential, offer high precision but require additional heating mechanisms. In comparison, Langmuir swept probes operate by continuously varying the applied voltage to measure the full characteristic $I$--$V$ curve of the plasma. This method allows for the comprehensive determination of plasma parameters, providing a more complete plasma characterization compared to other probe methods. However, swept probes are often challenged by higher-frequency fluctuations present in the plasma environment, introducing errors in the data. In addition, they require more complex electronic circuits for voltage control and data acquisition.

The growing interest in temporally resolving electron temperature, electron density, and plasma potential\cite{fast-swept-spatiotemporal-array, langmuir-low-ionization} stems from the need to understand fast transient phenomena in dynamic plasma environments, such as turbulence\cite{turbulence}, instability evolution\cite{hu2021probe}, pulsed plasma discharges\cite{pulsed-plasma} and confinements\cite{gao2022hot}. Typical Langmuir probes provide valuable time-averaged measurements, but lack the temporal resolution necessary to capture rapid fluctuations that significantly influence plasma behavior\cite{temporal-limits}. Fast-sweeping Langmuir probes address this limitation by enabling real-time tracking of plasma parameters on timescales from milliseconds to nanoseconds\cite{400khz}. These high-temporal resolution diagnostics allow for the identification of fast electron kinetics, instability growth, and wave-particle interactions, offering deeper insights into plasma transport and energy dissipation processes\cite{fluctuations-energy-transport, thomas1997fluctuation}. Their application in fusion research, space plasmas, and industrial plasma processing is increasingly vital to optimize performance and advance theoretical plasma models.

An example of these applications are direct current (DC) arc jet facilities, high-enthalpy plasma wind tunnels designed to simulate extreme aerodynamic heating conditions encountered during atmospheric entry, spacecraft re-entry, and hypersonic flight. These facilities generate high-enthalpy supersonic flows by sustaining a DC arc discharge between electrodes, heating a working gas to ionized states before accelerating it through a nozzle. The NASA Ames Arc Jet Complex is one of the most advanced facilities of this type, providing crucial testing environments for thermal protection systems (TPS) and ablative materials under realistic entry conditions~\cite{tsf_test_plan}. However, characterizing plasma properties in these environments remains a challenge due to their high enthalpy, non-equilibrium nature, and strong gradients in temperature and density~\cite{non-lte-arcjet, meurisse2018unsteady}. For example, there have been several studies in revealing the non-equilibrium nature of DC arc jets and its impact on numerical simulations, highlighting the importance of better understanding the plasma parameter behaviors~\cite{Guo2014, Baeva2016, ZubeMyers1993}.  Additionally, previous studies have shown low-frequency (1--10~kHz) oscillatory behaviors in the flow\cite{poulikakos1996electric, ahf-bdot-probes}, raising concerns about uncertainty in diagnostic measurements and modeling. Implementing Langmuir diagnostics in arc jet environments presents an opportunity to directly measure electron temperature, electron density, and plasma potential with high spatial and temporal resolution. These measurements are essential for improving computational models, understanding non-equilibrium plasma behavior, and refining heat transfer predictions, ultimately advancing the design and validation of TPS materials for future missions. \par 

Thus, the purpose of this paper is to design and develop a fast-sweeping Langmuir probe system that is open-source, low-cost, and adaptable for a wide range of plasma environments. By leveraging modern electronics and data acquisition techniques, the probe system aims to achieve fast voltage sweeping to resolve rapid fluctuations in electron temperature, electron density, and plasma potential. To validate its performance, the system is implemented and tested in the 30~kW DC miniature Arc jet Research Chamber (mARC~II), a small-scale arc jet facility developed to support the testing of novel and low technology readiness level concepts~\cite{Rodrigues2024, marc-2025-expanding, Marshall2025}. This facility provides an ideal testbed for assessing the probe’s capability to operate in high-enthalpy, non-equilibrium plasmas. The results will not only demonstrate the feasibility of using fast-sweeping Langmuir diagnostics in arc jet flows, but also provide an accessible and cost-effective solution for researchers studying transient plasma behavior in various applications.

 \vspace{-10pt}
%%%%%%%%%%%%%%%%%%%%%%%%%%%%%%%%%%%%%%%%%%%%%%%%%%%%%%%%%%%%%%%%%%%%%%%%%%%%%%%%%%%%
\section{Hardware Design}
%%%%%%%%%%%%%%%%%%%%%%%%%%%%%%%%%%%%%%%%%%%%%%%%%%%%%%%%%%%%%%%%%%%%%%%%%%%%%%%%%%%%
\vspace{-10pt}

The hardware design of the fast-sweeping Langmuir probe system is guided by key objectives that align with the broader goals for high-voltage ground testing facilities. These objectives include (1) ensuring a cost-effective solution to enhance accessibility, (2) implementing a battery-powered system to enable portability and electrical isolation from facility power supplies, reducing noise and interference, (3) providing adjustable frequency and voltage sweeping capabilities to accommodate a wide range of plasma conditions and measurement requirements, and (4) designing a robust probe capable of withstanding extreme aerothermal environments, such as those found in high-enthalpy arc jet flows. The following sections detail the circuit and hardware design to meet these performance criteria.

\vspace{-10pt}
%%%%%%%%%%%%%%%%%%%%%%%%%%%%%%%%%%%%%%%%%%%%%%%%%%%%%%%%%%%%%%%%%%%%%%%%%%%%%%%%%%%%
\subsection{Power Supply Array Board}
%%%%%%%%%%%%%%%%%%%%%%%%%%%%%%%%%%%%%%%%%%%%%%%%%%%%%%%%%%%%%%%%%%%%%%%%%%%%%%%%%%%%
\vspace{-10pt}

\begin{figure}[h]
\centering

\includegraphics[width=8.5cm, height=5.5cm]{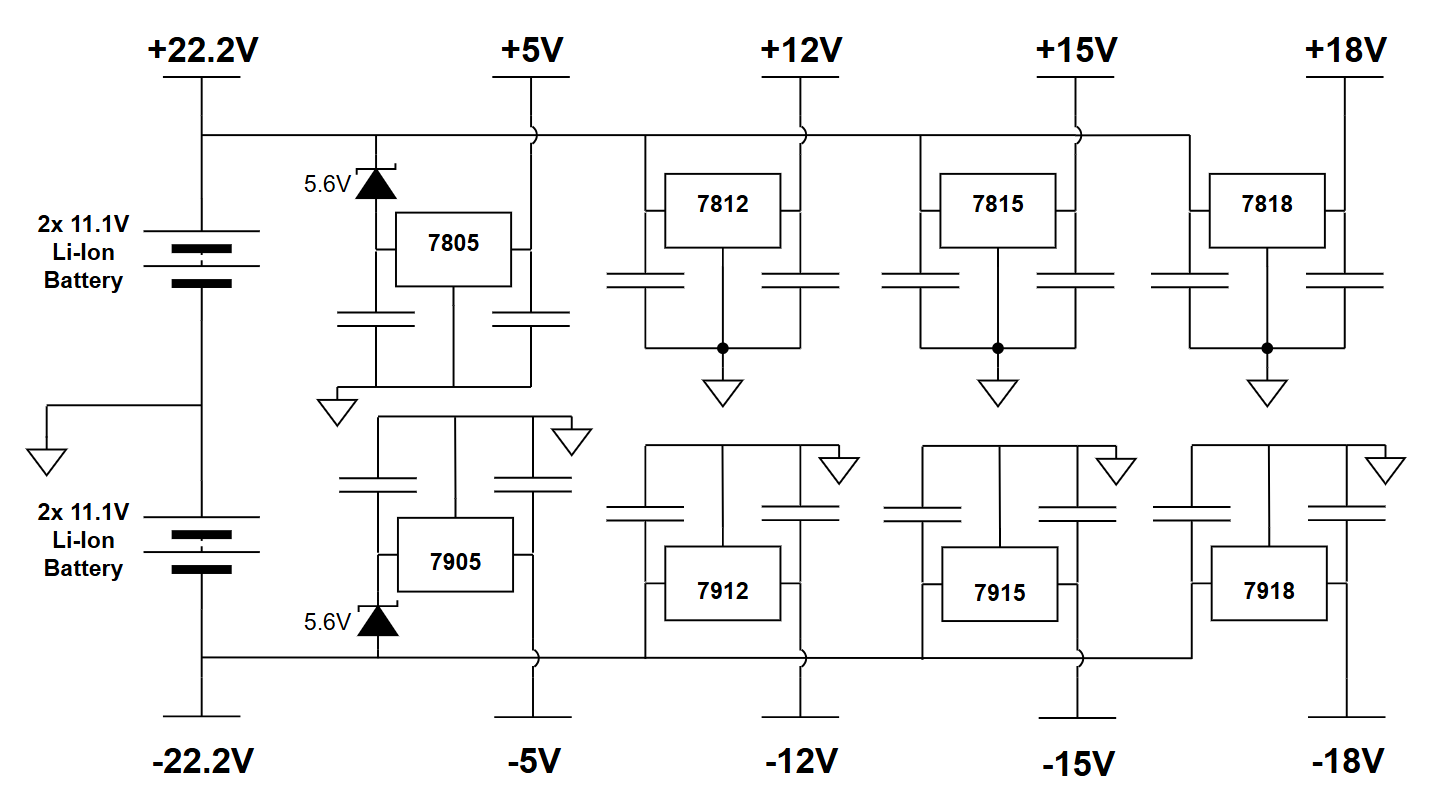}

\caption{Circuit schematic of power supply array. Note that a high-power zener diode was placed in series with the input of the $\pm$5~V regulators to create a 5.6~V drop or, in other words, bring the input voltage to an acceptable range as according to the datasheet.}
\label{fig:power-supply-array}
\end{figure}

As shown in Fig.~\ref{fig:power-supply-array}, the power supply uses four 11.1~V Li-Ion batteries to create $\pm$22.2~V rails. These voltage lines are then fed into a series of linear voltage regulators of the 78XX and 79XX family to establish $\pm$5~V, $\pm$12~V, $\pm$15~V, and $\pm$18~V. Each of these regulators can supply over 1.5~A with sufficient heat sinking. Additionally, they implement internal current limitation, thermal shutdown, and safe area compensation, making these robust and ideal for higher power applications~\cite{78XX}. 

\vspace{-10pt}
%%%%%%%%%%%%%%%%%%%%%%%%%%%%%%%%%%%%%%%%%%%%%%%%%%%%%%%%%%%%%%%%%%%%%%%%%%%%%%%%%%%%
\subsection{Voltage Multiplier Board}
%%%%%%%%%%%%%%%%%%%%%%%%%%%%%%%%%%%%%%%%%%%%%%%%%%%%%%%%%%%%%%%%%%%%%%%%%%%%%%%%%%%%
\vspace{-10pt}

\begin{figure*}
\centering

\includegraphics[width=1\textwidth]{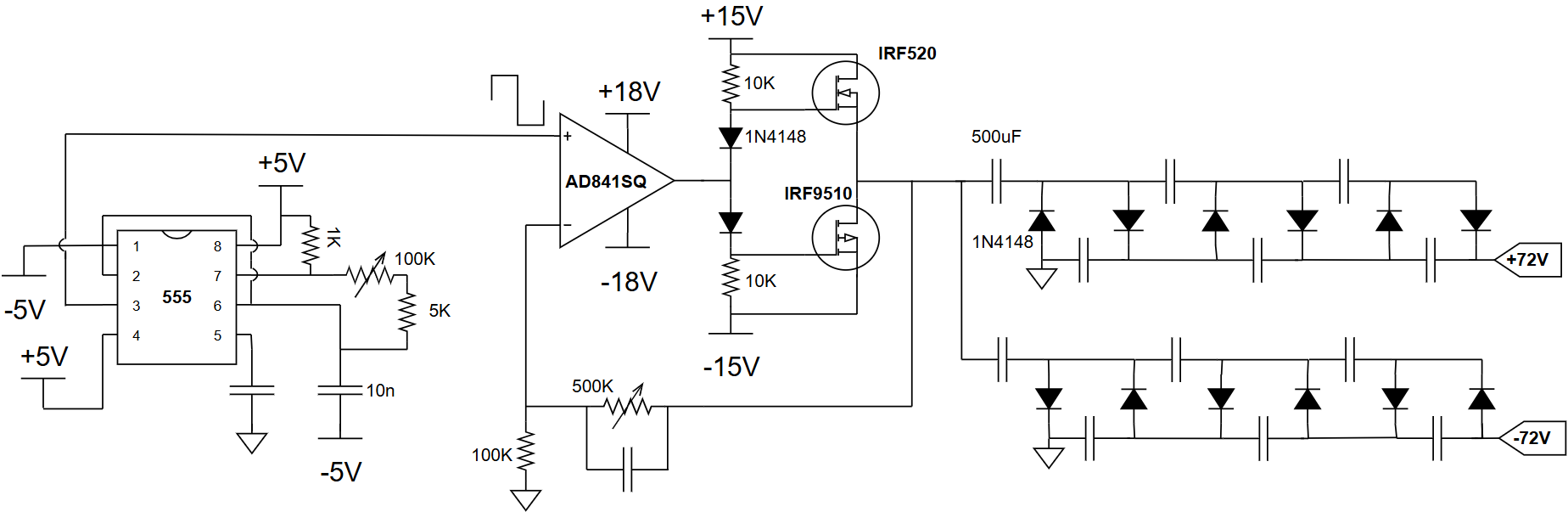}

\caption{Circuit schematic of voltage multiplier board. In this diagram, the board is configured to have three-stage multiplication, producing $\pm$72~V from 12~V AC signal.}
\label{fig:voltage-board}
\end{figure*}

The purpose of the voltage multiplier board is to establish high-voltage rails between which the driven signal can swing. As illustrated in Fig.~\ref{fig:voltage-board}, the board implements an oscillator and a cascade of voltage multipliers to convert a small AC voltage to a high DC voltage. The oscillator is a 555-timer circuit that outputs a 10~Vpp square wave that has an adjustable output frequency that ranges from 800~Hz to 50~kHz. This square-wave signal is fed into a non-inverting amplifier constructed from an AD841 op-amp. This op-amp was selected because of its high slew rate, which minimizes the distortion of higher-frequency signals. This op-amp amplifies the square-wave signal to 24~Vpp and is then fed into a power amplifier constructed from a couple of high-power MOSFETs (i.e., IRF520 and IRF9510). The purpose of the power amplifier is to supply a stable, high-current AC signal to the voltage multiplier. This ensures a stable current supply to the high-value capacitors and prevents a significant voltage drop when the voltage multiplier is under load. \par  

The voltage multiplier circuit implements high-speed switching 1N4148 diodes and 500~$\upmu$F capacitors. The high capacitance increases the amount of holding charge, and thus increases the amount of current that can be supplied. The number of voltage multiplication stages can be configured to be three or four, providing either $\pm$72~VDC or $\pm$96~VDC, respectively. This allows the user to expand to high voltages if necessary, with the caveat that it is more difficult for the circuit to supply higher currents as a result of the extra stage. 

\vspace{-10pt}
%%%%%%%%%%%%%%%%%%%%%%%%%%%%%%%%%%%%%%%%%%%%%%%%%%%%%%%%%%%%%%%%%%%%%%%%%%%%%%%%%%%%
\subsection{Signal Amplifier Board}
%%%%%%%%%%%%%%%%%%%%%%%%%%%%%%%%%%%%%%%%%%%%%%%%%%%%%%%%%%%%%%%%%%%%%%%%%%%%%%%%%%%%
\vspace{-10pt}

\begin{figure*}
\centering
\includegraphics[scale=0.4]{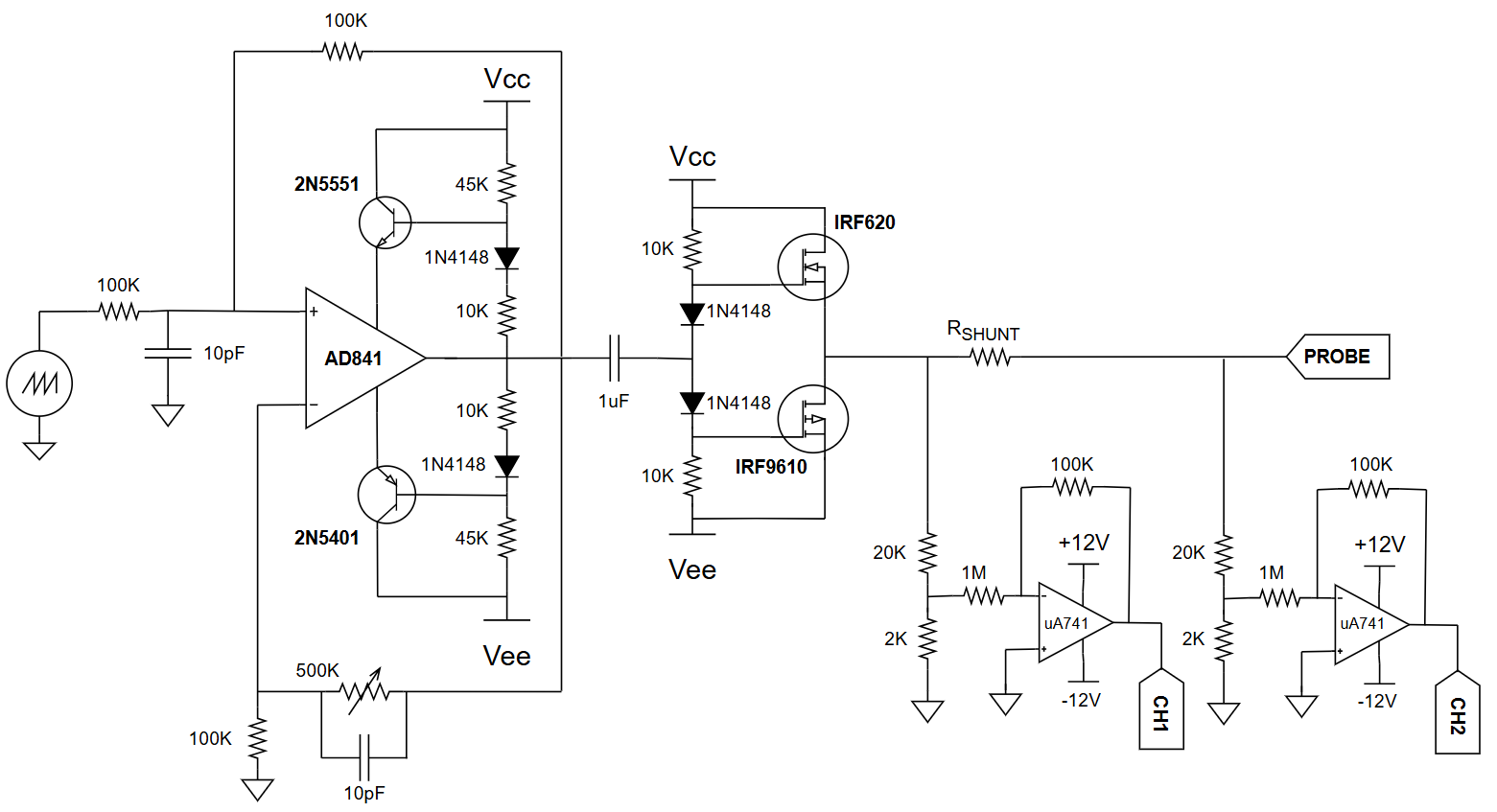}
\caption{Circuit schematic of signal amplifier board.}
\label{fig:signal-amplifier-board}
\end{figure*}

The signal driver circuit implements a boot-strap amplifier and a push-pull amplifier to amplify a sweeping signal to high voltages with high-current sourcing/sinking necessary for Langmuir applications. As shown in Fig.~\ref{fig:signal-amplifier-board}, the bootstrap amplifier is constructed from a high slew-rate AD841 op-amp, high-power 2N5551~NPN transistor, high-power 2N5401~PNP transistor, and other passive components. The transistors control the voltage across the op-amp's supply lines, swinging it to follow the high-voltage output. This allows the supply lines of the op-amp to be within an acceptable range, while amplifying the input signal to near-DC voltage produced by the voltage multiplier. The op-amp uses a non-inverting configuration, and the gain of amplification can be adjusted through a potentiometer. The output is fed into a push-amplifier constructed from high-power MOSFETs (i.e., IRF620 and IRF9610) that can handle voltages up to 200~V and high-current sourcing/sinking (up to 1.6~A)\cite{vishay_irf620, vishay_irf9610}. In the end, the board amplifies the input signal to a high-voltage signal of up to 120~Vpp, while sourcing/sinking sufficient current for Langmuir applications. \par 

The board additionally integrates a shunt resistor in series with the output, which is used to measure the current for which the Langmuir probe is sourcing or sinking. The resistance can be configured to be 100~$\Omega$, 200~$\Omega$, 270~$\Omega$, or 470~$\Omega$. The selection of the series resistance in a way changes the sensitivity to the amount of current measured. For example, for a 100 $\Omega$ resistor, a current of 10~mA produce a voltage of 1~V across the resistor. Meanwhile, a 470~$\Omega$ resistor produces a delta of 4.7~V. Thus, it is important to select the resistor carefully during application. \par 

However, many oscilloscopes or other data acquisitions have a severe bandwidth limitation when measuring high-voltage signals. Thus, the voltage before and after the shunt resistor is fed into a voltage divider and an inverting op-amp. This attenuates the signal to an acceptable range so that the oscilloscope can acquire data at high acquisition rates. 

\vspace{-10pt}
%%%%%%%%%%%%%%%%%%%%%%%%%%%%%%%%%%%%%%%%%%%%%%%%%%%%%%%%%%%%%%%%%%%%%%%%%%%%%%%%%%%%
 \subsection{Signal Generator Board}
%%%%%%%%%%%%%%%%%%%%%%%%%%%%%%%%%%%%%%%%%%%%%%%%%%%%%%%%%%%%%%%%%%%%%%%%%%%%%%%%%%%%
\vspace{-10pt}

\begin{figure}
\centering

\includegraphics[width=8.5cm, height=4cm]{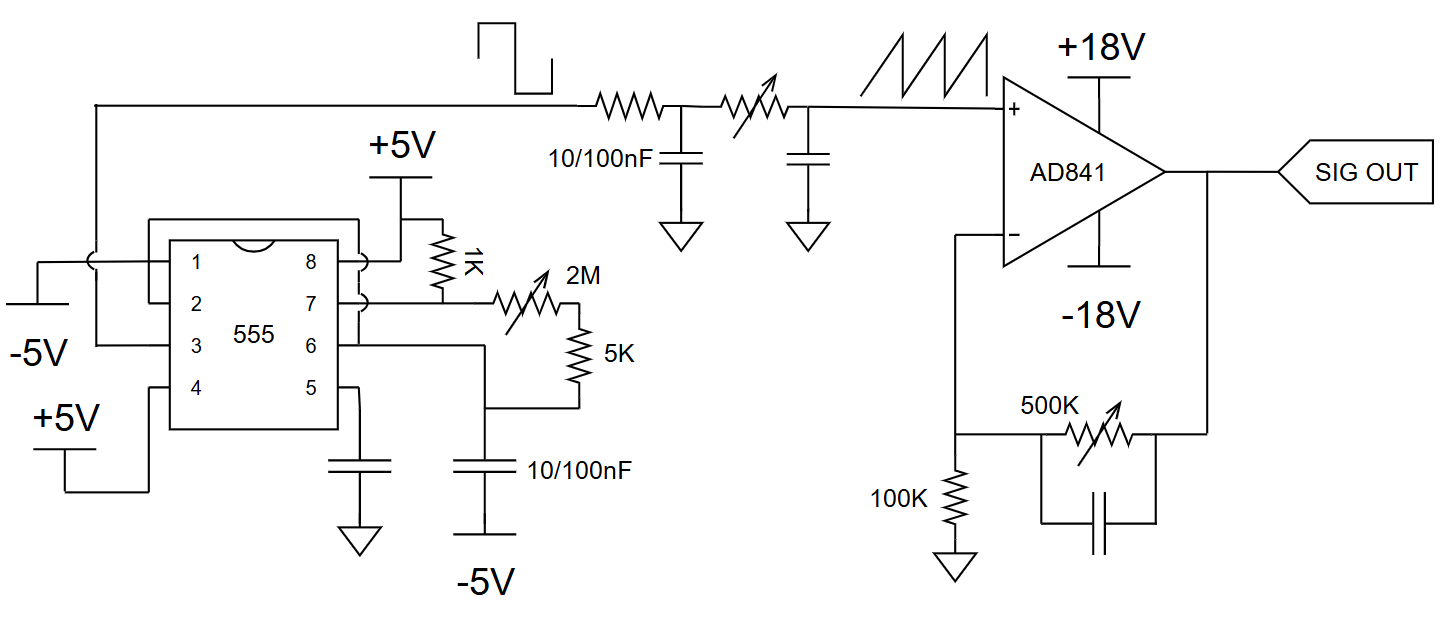}

\caption{Circuit schematic of signal generation board.}
\label{fig:signal-generation-board}
\end{figure}

The motivation behind the signal generation board is to minimize the need for a bench-top signal generator, and thus minimize ground loops, reducing the susceptibility to EMI noise. As illustrated in Fig.~\ref{fig:signal-generation-board}, the first stage of the board is a 555-timer circuit that produces a 5~V amplitude square-wave signal with an adjustable frequency, ranging from 300~Hz to 100~kHz. The square-wave signal is then fed to a variable two-stage low-pass RC filter to produce a sawtooth wave. The filtered signal is last amplified by an inverting AD841 op-amp circuit with adjustable gain, capable of outputting amplitudes of up to 15~V. The AD841 offers full power bandwidth with 200~k$\Omega$ output impedance, ensuring minimal signal distortion when under significant current loads. 

\vspace{-10pt}
%%%%%%%%%%%%%%%%%%%%%%%%%%%%%%%%%%%%%%%%%%%%%%%%%%%%%%%%%%%%%%%%%%%%%%%%%%%%%%%%%%%%
\subsection{Probe}
%%%%%%%%%%%%%%%%%%%%%%%%%%%%%%%%%%%%%%%%%%%%%%%%%%%%%%%%%%%%%%%%%%%%%%%%%%%%%%%%%%%%
\vspace{-10pt}

The probe was constructed from 16~AWG magnet wire, where the insulation was carefully stripped to create a cylindrical probe of approximately 1~mm length and 1.3~mm diameter, or 1.3~$\text{mm}^2$ probe area. The probe tip was sanded to remove sharp geometries and thus avoid non-uniform Debye sheaths\cite{huddlestone1965langmuir}. The probe, except for the exposed area, was wrapped in Kapton and fiberglass masking tape, providing additional thermal, electrical, and abrasive protection from the extreme aerothermal loads inside the mARC~II test box (see~\S\ref{sec:marc}. Implementation: mARC~II Facility). \par 
Due to the high densities of plasmas typically found in arc jet environments, the Debye length is small (discussed further in ~\S\ref{results}.~Results) meaning the collection area of the probe is approximately the probe area. As such, we estimate the spatial resolution of the probe to be approximately the length scale of the probe ($\sim$~1~mm).

\vspace{-10pt}
%%%%%%%%%%%%%%%%%%%%%%%%%%%%%%%%%%%%%%%%%%%%%%%%%%%%%%%%%%%%%%%%%%%%%%%%%%%%%%%%%%%%
\section{Preliminary Testing}
\label{sec:preliminary-testing}
%%%%%%%%%%%%%%%%%%%%%%%%%%%%%%%%%%%%%%%%%%%%%%%%%%%%%%%%%%%%%%%%%%%%%%%%%%%%%%%%%%%%
\vspace{-10pt}

Prior to implementation, it is important to characterize the system's performance. This includes measuring the half-gain bandwidth, gain range, max voltage swing, and non-linearity. The results are shown in Table~\ref{table:langmuir-properties}. 

\begin{table}[b]
\caption{Electrical properties of the in-house developed fast-sweep Langmuir probe prototype.}
\begin{ruledtabular}
\begin{tabular}{cc}
\vspace{1mm}
Features&Value\\
\hline
\vspace{0.5mm}
Half-gain bandwidth&200~kHz\\
\vspace{0.5mm}
Gain range&1--5\\
\vspace{0.5mm}
Voltage range (max)&$\pm$60~V\\
\vspace{0.5mm}
Non-linearity&4$\%$\\
\end{tabular}
\end{ruledtabular}
\label{table:langmuir-properties}
\end{table}

\begin{table}[b]
\caption{\label{tab:limits} Application limits of the in-house developed fast-sweep Langmuir probe prototype.}
\begin{ruledtabular}
\begin{tabular}{cc}

\vspace{1mm}
Feature&Limit\\
\hline
\vspace{0.5mm}
Parameter Fluctuations (max)$^*$&100~kHz\\
\vspace{0.5mm}
Spatial Resolution (min)$^{**}$&1 mm\\
\vspace{0.5mm}
Electron velocity (max)$^{\dagger}$&$21\times10^6$ m/s\\
\vspace{0.5mm}
Electron density (min)$^{\dagger\dagger}$&$1\times10^{11}$ cm$^{-3}$
\end{tabular}
\begin{flushleft}\vspace{-5pt}
      \small
      $^*$Nyquist frequency of max sweeping frequency $f_{\rm{sweep (max)}}$.\par
      $^{**}$When debye length is significantly smaller than probe's length scale, spatial resolution is approximately probe's length scale.\par
      $^{\dagger}$Calculated using the max probe voltage and energy conservation. \par 
      $^{\dagger\dagger}$Estimated using $T_e = 10$ eV and $R_p > 10\times\lambda_D$ for thin-sheath model to be applicable. \par 
\end{flushleft}
\end{ruledtabular}
\label{table:limits}

\end{table}

\begin{figure}
\centering
\includegraphics[width=9cm]{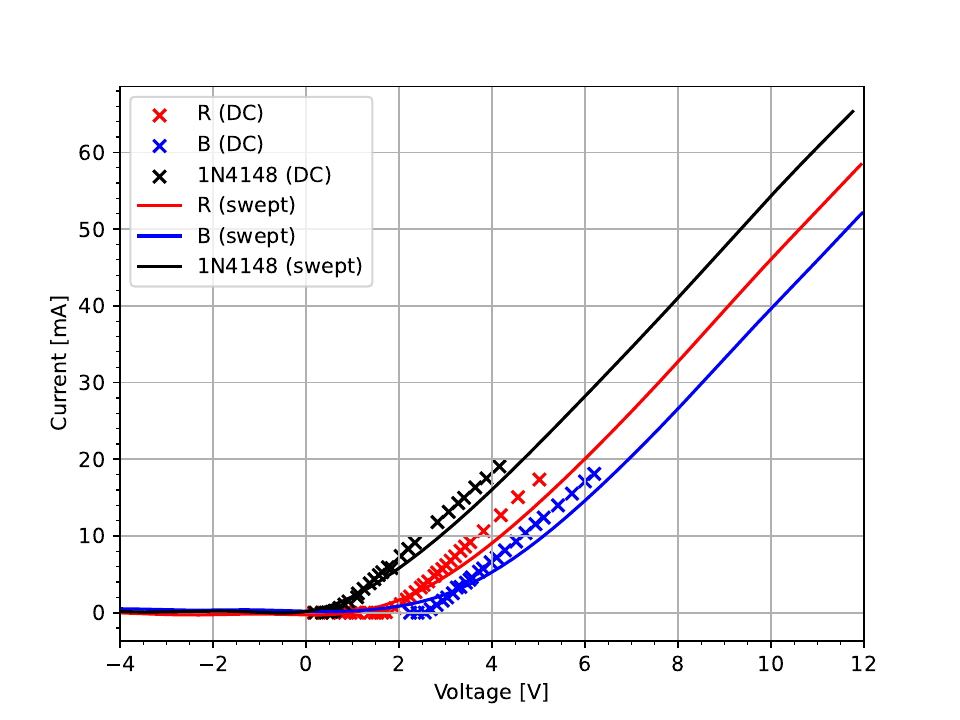}
\caption{Characteristic $I$--$V$ curve of 1N4148 rectifier diode~(\solblack), red LED~(\solred), and blue LED~(\solblueone) measured by using the fast-sweeping Langmuir probe and manually measuring voltage drop across the shunt resistor at different applied DC input voltages~($\boldsymbol{\times}$, \textcolor{red}{$\boldsymbol{\times}$}, \textcolor{blue}{$\boldsymbol{\times}$}).}
\label{fig:IV-curve-diode}
\end{figure}

The functionality of the circuit was verified by driving different diodes and measuring the characteristic $I$--$V$ curve. The diodes include fast-switching diodes, rectifier diodes, and light-emitting diodes (LEDs). The diodes were driven with a 50~kHz, 20~Vpp signal and the resistance was measured with a 170~$\Omega$ shunt resistor. Figure~\ref{fig:IV-curve-diode} illustrates the characteristic $I$--$V$ curves for a 1N4148 fast-switching diode, a red LED, and a blue LED. It is important to note that this current design of the Langmuir probe system is uncompensated and therefore prone to stray capacitance. To counteract this issue, the stray capacitance was measured by obtaining the current through the shunt resistor when no load is applied and, after averaging,  subtracted from the raw data. These curves were manually compared to the $I$--$V$ curves measured with an applied DC voltage. The comparison between the two techniques shows similarity and verifies the functionality of the system. The forward voltage was estimated by fitting a line to the current when the diode is conducting and measuring the line's voltage-intercept. This simulates the measurement of the floating potential in a Langmuir $I$--$V$ curve and in this test showed an average accuracy of 10.9$\%$. Furthermore, shunt resistance was estimated by measuring the slope of the linear $I$--$V$ curve, simulating the measurement of the electron temperature, and showed an average accuracy of 3.2$\%$. This level of accuracy is competitive with other Langmuir probe designs~\cite{400khz, chen2016introduction} and, more importantly, proves to be sufficient for measuring plasma parameters. \par

Additionally, it is beneficial to estimate the functionality of this diagnostic for different regimes of plasmas. Table~\ref{table:limits} showcases some limitations in parameter features that can be measured using this diagnostic. 

\vspace{-10pt}
%%%%%%%%%%%%%%%%%%%%%%%%%%%%%%%%%%%%%%%%%%%%%%%%%%%%%%%%%%%%%%%%%%%%%%%%%%%%%%%%%%%
\section{Implementation: \MakeLowercase{m}ARC~II Facility}
\label{sec:marc}
%%%%%%%%%%%%%%%%%%%%%%%%%%%%%%%%%%%%%%%%%%%%%%%%%%%%%%%%%%%%%%%%%%%%%%%%%%%%%%%%%%%%
\vspace{-10pt}

The mARC~II is an in-house testing facility aimed at supporting lower technology readiness level~(TRL) technologies for reentry applications. The facility revolves around its 30~kW segmented constrictor-type arc heater consisting of a cathode, two or three constrictor disks (one integrated with a pressure tap), an anode, and a converging-diverging nozzle. The power supply and compressed air flow is controlled by the Hypertherm MAX200 plasma cutter, capable of 200 A current supply and 1.0~g/s gas flow. The vacuum system utilizes an Edwards EH4200 booster pump and an Edwards E2M275 rotary vane pump, causing the test box pressures to reach a stable pressure of approximately 2~Pa during steady-state operation~\cite{marc-2024-defining, marc-2025-expanding}. Due to this pressure and power range, the facility produces a collision-dominated, low-ionization plasma jet with bulk enthalpies ranging from 3 to 22~MJ/kg. \par 
%reaching up to 22~MJ/kg bulk flow enthalpy. \par 

\begin{figure}
\centering
\includegraphics[width=9cm]{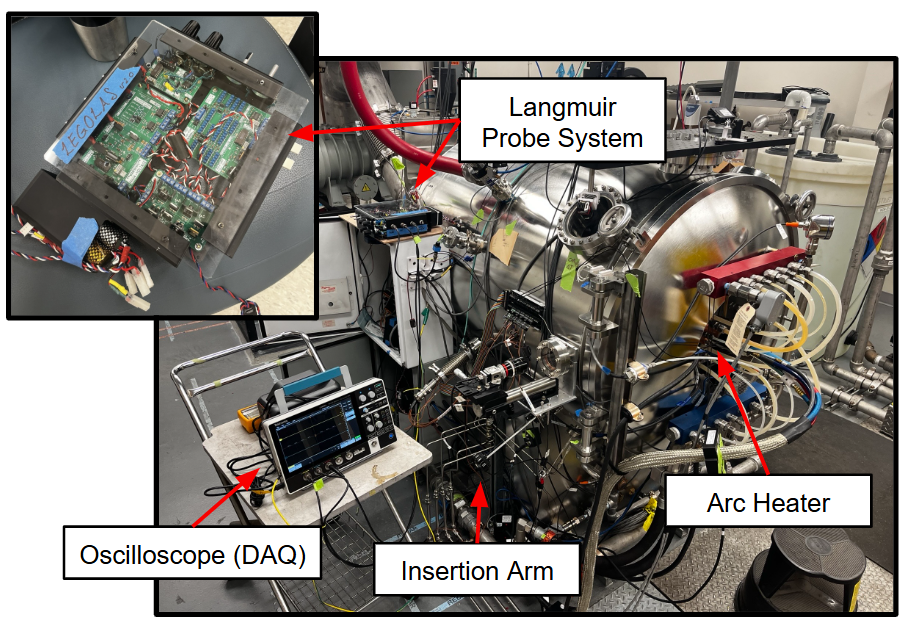}
\caption{Arc jet facility: mARC~II test box highlighting the arc heater, Langmuir probe system, insertion arm, and data acquisition.}
\label{fig:mARC}
\end{figure}

The Langmuir probe is inserted into the flow using a rotary linear direct feed-through. The insertion arm is coupled with a potentiometer-based encoder to provide the radial distance of the probe from the flow centerline. The probe sits approximately 50~mm from nozzle exit when fully inserted. Due to limitations of the facility's test series, the axial distance could not be adjusted. Although it is possible to measure closer to the nozzle exit, one has to be wary of the high-enthalpy environment, thus taking advantage of the fast insertion capability of this probe. The data is acquired using a digital oscilloscope and is configured to sample at 50~MHz (with a total acquisition duration of 0.2~s) when triggered by the falling edge of the encoder signal. This sample duration was sufficient to capture the fast insertion of the probe into the flow ($\sim$~0.1~s) and acquire sufficient data points ($\sim$~500~points) for every voltage sweep. For users implementing similar technology, it is suggested to select a sufficiently high acquisition rate such that there are sufficient data points at each sweep to capture rapidly changing response (i.e. $f_{\rm acq} \geq 10\times f_{\rm sweep}$) while capturing the full insertion. Figure~\ref{fig:mARC} illustrates the facility's test box and the Langmuir probe system setup. \par

\begin{figure}
\centering
\includegraphics[width=9cm]{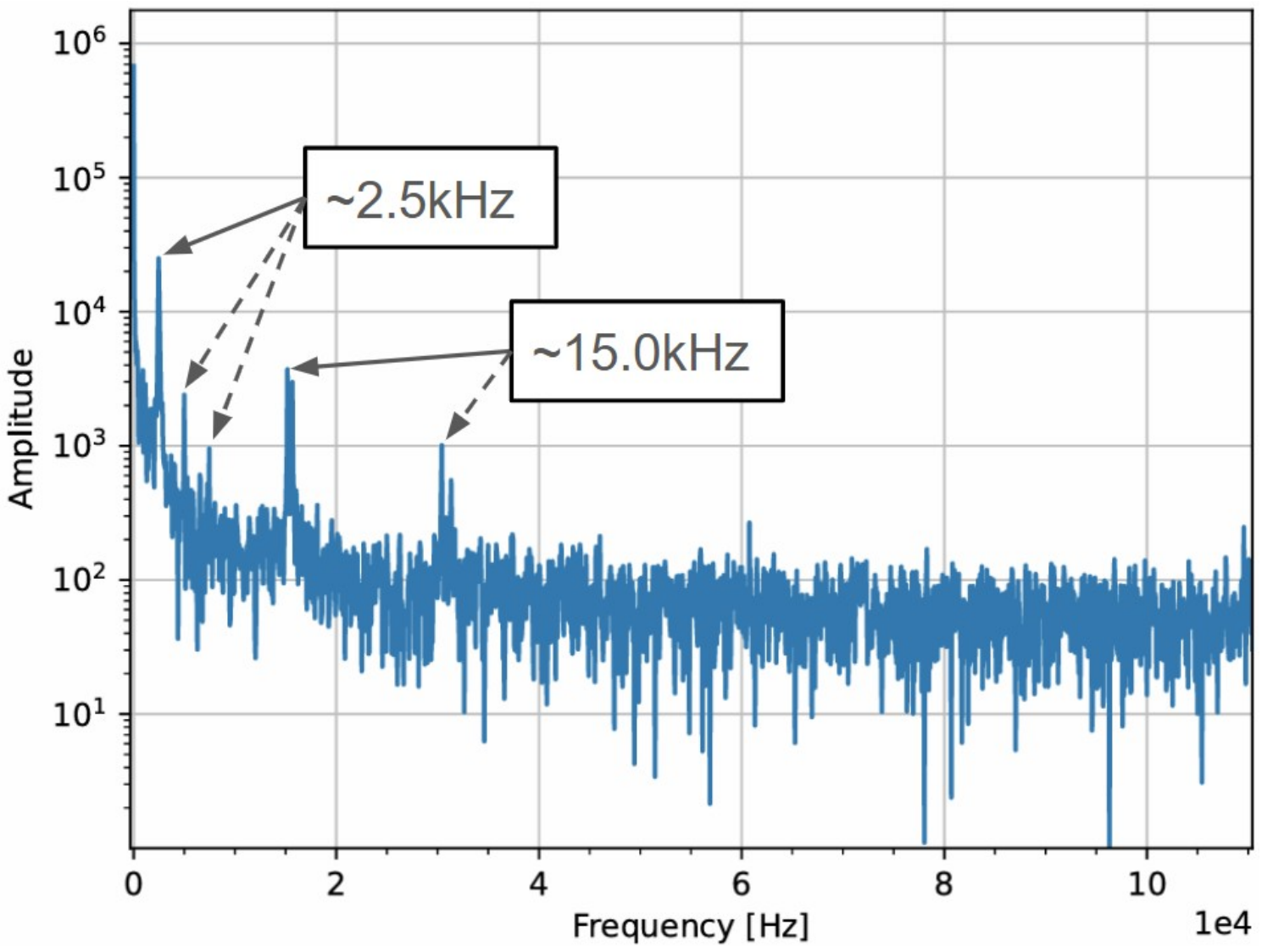}
\caption{Frequency spectrum of the flow obtained by inserting a floating probe at a distance of 50~mm from the nozzle exit. The conditions for this test 200~A current and gas flow rate of 0.35~g/s. The prominent frequencies $\sim$2.5~kHz and $\sim$15~kHz are indicated~(\grayarrow) as well as their respective harmonics~(~\graydashedarrow).}
\label{fig:frequency-spectrum}
\end{figure}

Before the test, the stray capacitance of the system was measured similarly to what was conducted in \S\ref{sec:preliminary-testing}.~Preliminary Testing. In one test, the probe tip was directly connected to the oscilloscope in the motivation to measure the fluctuation behavior in the plasma flow. Figure~\ref{fig:frequency-spectrum} shows the frequency spectrum (after the background scan is subtracted). As seen in the data, there are broad low-frequency fluctuations present within the flow, making a 100~kHz sweeping frequency sufficient for temporally obtaining plasma parameters. There are notable peaks at $\sim$2.5~kHz and $\sim$15~kHz, including their respective harmonics. The origin of these modes is still unknown and is currently attributed to the oscillatory behavior of the arc between the electrodes\cite{poulikakos1996electric}. Future efforts will aim to better understand and characterize the frequency behaviors present within the plasma jet. \par 

% Need to clarify the origin of the 15kHz - Need to read the user manual. Neet to mention that this frequency peak occurs in multiple runs. 

The sweeping frequency was configured to be 100~kHz with a voltage range of $-$40~V to 40~V. The probe is inserted after the plasma jet has reached steady-state condition.

\vspace{-10pt}
%%%%%%%%%%%%%%%%%%%%%%%%%%%%%%%%%%%%%%%%%%%%%%%%%%%%%%%%%%%%%%%%%%%%%%%%%%%%%%%%%%%%
\section{Results}
\label{results}
%%%%%%%%%%%%%%%%%%%%%%%%%%%%%%%%%%%%%%%%%%%%%%%%%%%%%%%%%%%%%%%%%%%%%%%%%%%%%%%%%%%%
\vspace{-10pt}

The following subsections study how deviations from Maxwellian distributions impact current-voltage response and, using orbital motion-limited (OML) theory, electron temperature measurements. Afterwards, the experimental data obtained from the mARC~II test series and data analysis is presented. Prior to that, it is important to estimate the fundamental plasma parameters to determine the plasma regime for which this study resides and to provide a better understanding of the analysis and results~\cite{bellan2006fundamentals}. Using the findings of similar studies, the plasma density and temperature were chosen to be $1.0~\times~10^{14}$~cm$^{-3}$ and 10~eV, respectively~\cite{mARC-langmuir, johnson1997electron, Diana2021}. From this, the plasma parameters were estimated (shown in Table~\ref{table:plasma-parameters}) and used for the following subsection.

\vspace{-10pt}

\begin{table}[b]
\caption{Plasma parameters estimated using electron density $n_e$~=~$1.0~\times~10^{14}$~cm$^{-3}$ and electron temperature $T_{eV}$~=~10~eV.}
\begin{ruledtabular}
\begin{tabular}{cc}
\vspace{1mm}
Features&Value\\
\hline
\vspace{0.5mm}
Electron plasma frequency&$8.97 \times 10^{10}$~Hz\\
\vspace{0.5mm}
Ion plasma frequency&$2.10 \times 10^{9}$~Hz\\
\vspace{0.5mm}
Debye length& $2.28 \times 10^{-6}$ m\\
\vspace{0.5mm}
Electron thermal velocity& $1.32 \times 10^{7}$ m/s\\
\vspace{0.5mm}
Electron gyrofrequency& $2.80 \times 10^{5}$~Hz\\
\end{tabular}
\end{ruledtabular}
\label{table:plasma-parameters}
\end{table}

%%%%%%%%%%%%%%%%%%%%%%%%%%%%%%%%%%%%%%%%%%%%%%%%%%%%%%%%%%%%%%%%%%%%%%%%%%%%%%%%%%%%
\subsection{Model Sensitivity Analysis}
%%%%%%%%%%%%%%%%%%%%%%%%%%%%%%%%%%%%%%%%%%%%%%%%%%%%%%%%%%%%%%%%%%%%%%%%%%%%%%%%%%%%
\vspace{-10pt}

It is important to compare Langmuir's OML theory with other theories, such as the Allen-Boyd-Reynolds (ABR) theory and the Bernstein-Rabinowitz-Laframboise (BRL) theory. In particular, OML theory describes the current collection by a small probe in a plasma, assuming that charged particles follow ballistic trajectories influenced by the probe's potential, with negligible collisional or sheath effects\cite{chen2016introduction}. Due to the higher pressure conditions in arc jet environments, the electron density is high and thus the Debye length is small relative to low-density plasmas. Thus, the normalized probe length $\xi$~=~$\frac{r_p}{\lambda_D} \sim 400$. Studies conducted by~\citet{chen2016introduction}, while varying $\xi$, compared the results of these theories for high-density RF plasmas\cite{chen2016introduction, chen2020lecture}. The results show that for higher values of $\xi$ the electron temperatures agree across OML, ABR, and BRL. However, for electron density, there can be significant discrepancies, specifically BRL always reporting significantly higher values, while ABR and OML converge to similar values. This is due to the fact that ABR theory does not take orbital motions into account. Thus, in a collision-dominated environment where orbital motions are constricted due to short mean-free path, OML theory converges towards ABR theory. On the other hand, the BRL theory includes the orbital motion from ions far from the probe and thus underpredicts the ion-saturation current, consequently estimating significantly higher densities. Although this provides some insight, there is still a lack of understanding of how these theories perform for low-ionization, collisional plasmas. This motivates future efforts in small-scale numerical simulations to develop a model that is sensitive to non-Maxwellian distributions. \par

There is still debate as to whether the plasma produced in arc jet facilities' freestream flows follow thermal equilibrium or Maxwellian distributions due to their highly collisional and low-ionization nature~\cite{non-lte-arcjet}. However, that debate is outside of the scope of this paper. Since our paper's objective is to verify the new Langmuir probe diagnostic, the data is analyzed using OML~theory\cite{huddlestone1965langmuir, chen2016introduction} which assumes a Maxwellian distribution. However, it is important to quantify the sensitivity of OML~theory towards non-Maxwellian plasmas, providing insight into the variability for the results presented in the following section. \par

\vspace{-10pt}
%%%%%%%%%%%%%%%%%%%%%%%%%%%%%%%%%%%%%%%%%%%%%%%%%%%%%%%%%%%%%%%%%%%%%%%%%%%%%%%%%%%%
\subsubsection{Maxwellian Distribution with Suprathermal Beam}
\label{sec:maxwellian}
%%%%%%%%%%%%%%%%%%%%%%%%%%%%%%%%%%%%%%%%%%%%%%%%%%%%%%%%%%%%%%%%%%%%%%%%%%%%%%%%%%%%

To accomplish this, a 1D Maxwellian distribution combined with a suprathermal beam, a population of particles that have energies significantly higher than the thermal energy of the bulk plasma, was implemented. This was compared with a Maxwellian distribution model to study how the characteristic $I$--$V$ curve varied. Assuming thermal equilibrium and in the 1D case, the electrons follow a Maxwellian probability distribution function, which is defined as:
\begin{equation}
\label{eq:1D-Maxwellian-pdf}
    f(v) = \sqrt{\frac{m_e}{2 \pi k_B T_e}}e^{-\frac{m_ev^2}{2k_BT_e}}~.
\end{equation}

Here $m_e$ is the electron mass, $k_B$ is the Boltzmann constant, $T_e$ is the electron temperature, and $v$ is the particle velocity. Using this, one can calculate the current that the probe absorbs as given by:
\begin{equation}
\label{eq:current-1D-Maxwellian-pdf}
    I(V) = q_eA_p\Phi(V) = q_eA_p [n \int_{0}^{V} vf(v)dv]~.
\end{equation}

\begin{figure}
\centering
\includegraphics[width=8.5cm]{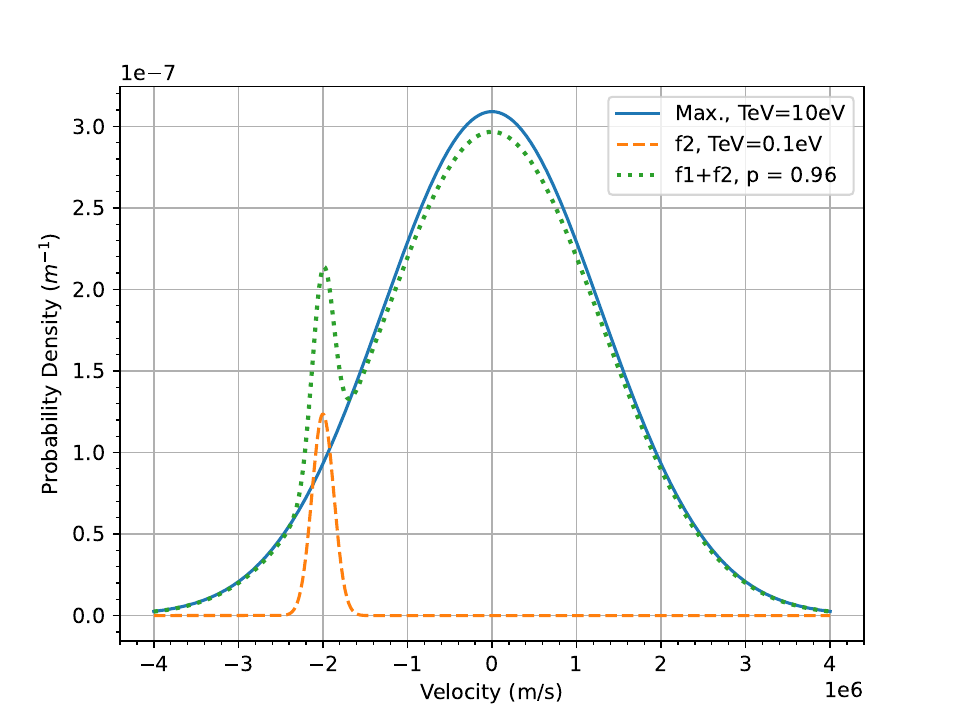}
\caption{A 1D suprathermal-Maxwellian electron probability density function, where $A_p$~$=$~$1$~mm$^{2}$, $n_e$~=~1~$\times$~$10^{14}$~$\text{cm}^{-3}$, $T_{eV_1}$~=~10~eV, $T_{eV_2}$~=~1~eV, and $v'$~=~2$\times10^6$~m/s. The orange-dashed~(\dashorange) and green-dotted~(\dotgreen) lines represents the suprathermal beam of electrons and the summation of these electrons with a Maxwellian distribution, respectively. The blue-line~(\solblue) represents the typical Maxwellian distribution with same plasma parameters.}
\label{fig:maxwellian-PDF}
\end{figure}

Here $q_e$ is the electron charge, $A_p$ is the area of the probe, $\Phi(V)$ is the electron density flux, and $V$ is the voltage that is applied to the probe. The suprathermal-Maxwellian distribution function is configured to have a primary population of electrons at temperature $T_{eV_1}$ and another population at $T_{eV_2}$ with a thermal-velocity offset of $v'$. Note that the probability density function needs to add to unity, thus:
\begin{equation}
\label{eq:f1-1D-Maxwellian-pdf}
    f_1(v) = p \sqrt{\frac{m_e}{2 \pi k_B T_{eV_1}}}e^{-\frac{m_ev^2}{2k_BT_{eV_1}}}~,
\end{equation}
\begin{equation}
\label{eq:f2-1D-Maxwellian-pdf}
    f_2(v) = (1 - p) \sqrt{\frac{m_e}{2 \pi k_B T_{eV_2}}}e^{-\frac{m_e(v - v')^2}{2k_BT_{eV_2}}}~,
\end{equation}

\noindent where $0 \leq p \leq 1$. When $p$~=~1, then the model returns to a single-temperature Maxwellian distribution function. Figure~\ref{fig:maxwellian-PDF} presents the electron probability density function for $p = 0.96$. Note that in this study, the plasma potential is taken to be 0~V for simplicity and thus the mean electron velocity for the Maxwellian distribution is 0~m/s. 

\vspace{40 pt}

\begin{figure}
\centering
\includegraphics[width=8.5cm]{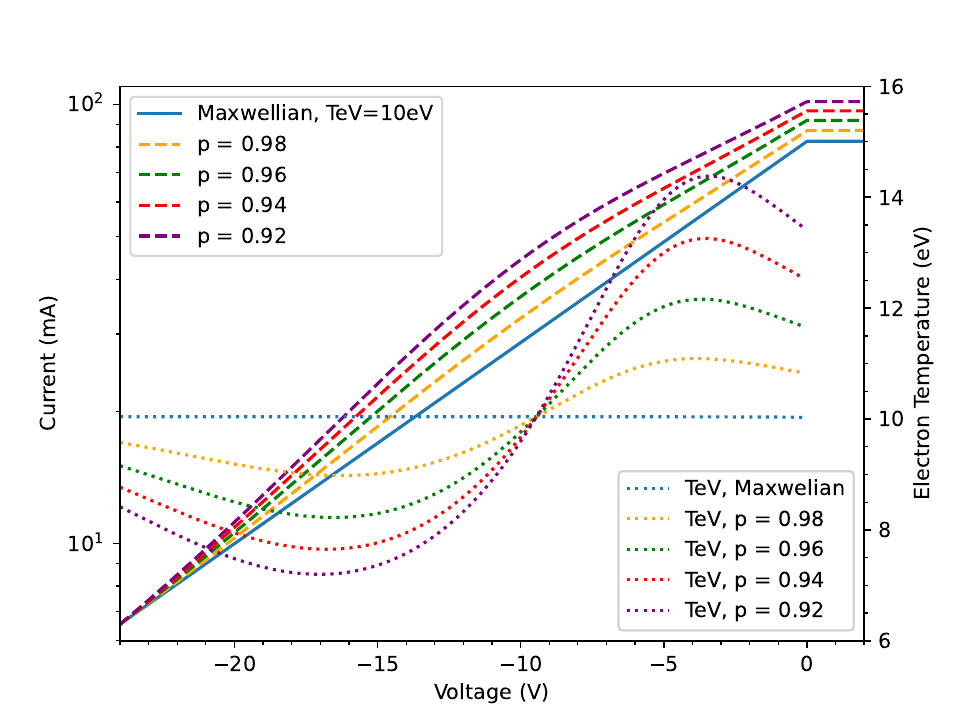}
\caption{Characteristic $I$--$V$ curve for the suprathermal-Maxwellian distribution model, while varying the $p$ parameter. The blue-line (\solblue) describes a single-temperature Maxwellian distribution, where $p$~=~1. The dotted lines represent electron temperature calculated from the $I$--$V$ curve using OML~theory.}
\label{fig:maxwellian-IV-curve}
\end{figure}

Using Eq.~\eqref{eq:current-1D-Maxwellian-pdf}, the electron characteristic $I$--$V$ curve was calculated and the results are shown in Fig.~\ref{fig:maxwellian-IV-curve}. In the the $-$24~V to $-$10~V region, there is an increase in current curve due to the suprathermal beam of electrons, causing deviation from the Maxwellian IV characterstic curve. This region corresponds to the lower tail ($\sim 2.4 \times 10^6$~m/s) and the peak ($2.0 \times 10^6$~m/s) of the suprathermal beam probability density function. However, the Maxwellian curve dominates the distribution function soon after causing convergence towards the Maxwellian curve. The suprathermal beam appears to largely affect the electron saturation current values and the probe current at the applied voltages. The electron temperature was measured from the $I$--$V$ curves using OML~theory. As seen in Fig.~\ref{fig:maxwellian-IV-curve}, the maximum deviation from the Maxwellian curve (occurring around $\sim16$~V) was calculated to be 11.9\%, 20.7\%, 28.7\%, and 33.8\% for $p$~=~0.98, 0.96, 0.94, and 0.92, respectively. Although the small initial change in $p$ shows an appreciable deviation, further increases in $p$ show a lesser effect. \par  

%%%%%%%%%%%%%%%%%%%%%%%%%%%%%%%%%%%%%%%%%%%%%%%%%%%%%%%%%%%%%%%%%%%%%%%%%%%%%%%%%%%%
\subsubsection{Two-Temperature Distribution}
%%%%%%%%%%%%%%%%%%%%%%%%%%%%%%%%%%%%%%%%%%%%%%%%%%%%%%%%%%%%%%%%%%%%%%%%%%%%%%%%%%%%

\begin{figure}
\centering
\includegraphics[width=8.5cm]{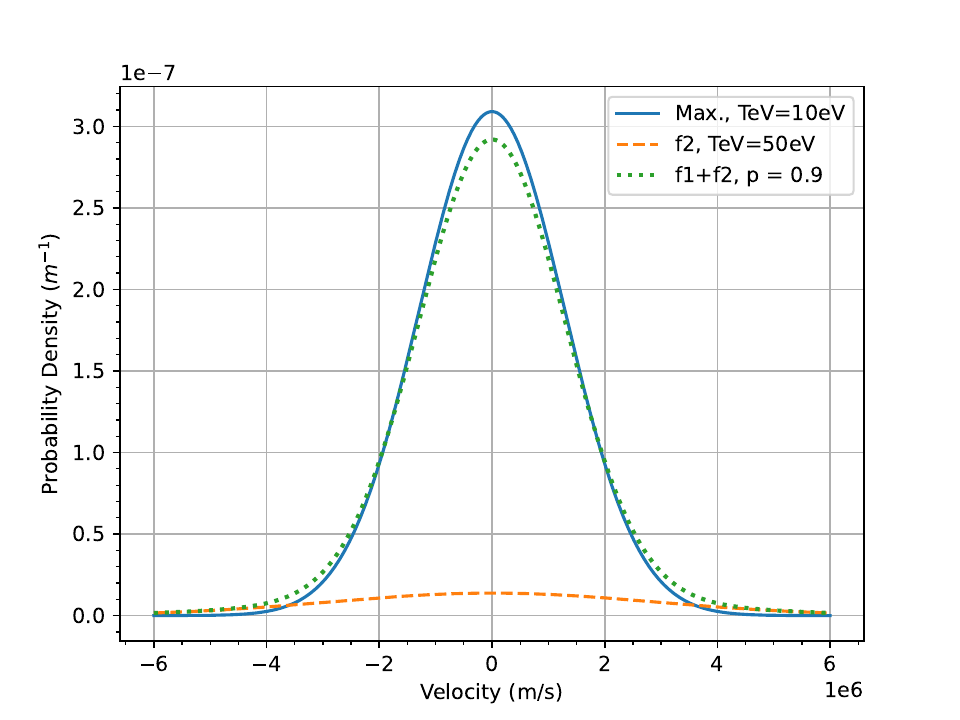}
\caption{A 1D two-temperature electron probability density function, where $A_p$~=~$1$~mm$^{2}$, $n_e$~=~1$\times10^{14}$~$\text{cm}^{-3}$, $T_{eV_1}$~=~10~eV, and $T_{eV_2}$~=~50~eV. The orange-dashed~(\dashorange) and green-dotted~(\dotgreen) lines represent the summation of the two temperature Maxwellian distributions. The blue-line~(\solblue) represents the typical Maxwellian distribution with same plasma parameters.}
\label{fig:two-temp-PDF}
\end{figure}

\begin{figure}
\centering
\includegraphics[width=8.5cm]{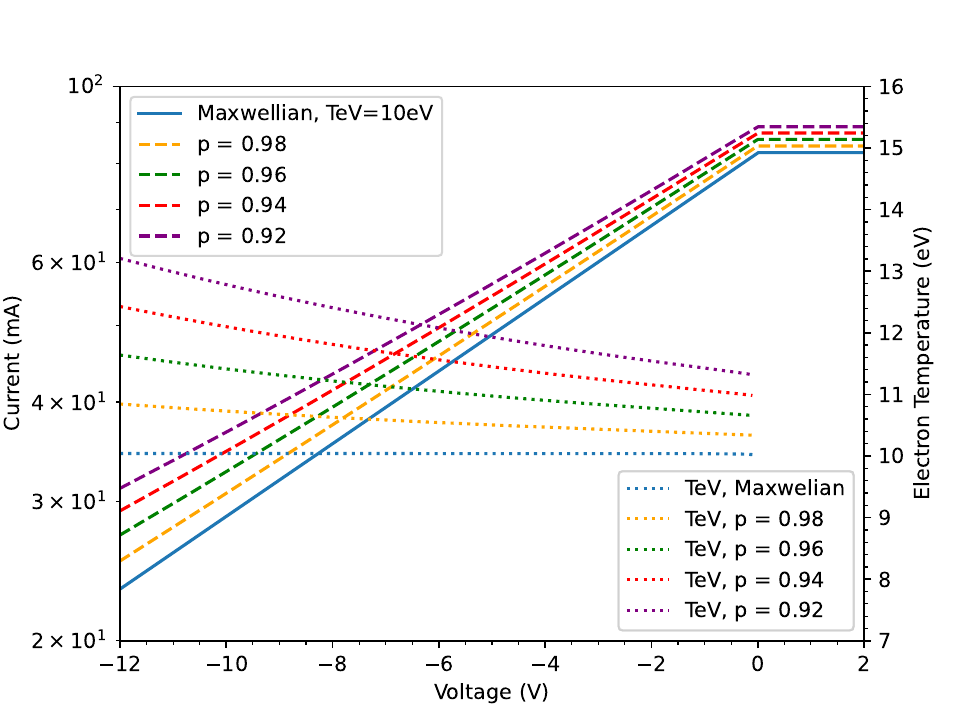}
\caption{Characteristic $I$--$V$ curve for the two-temperature distribution model, where $p$~=~1 describes a single-temperature Maxwellian distribution. The dotted lines represent electron temperature calculated from the $I$--$V$ curve using OML~theory.}
\label{fig:two-temp-IV-curve}
\end{figure}

The same analysis was conducted for a two-temperature model, where in Eq.~\eqref{eq:f2-1D-Maxwellian-pdf} the parameter $v'$ is set to 0~m/s. Figure~\ref{fig:two-temp-PDF} illustrates an example of the two-temperature distribution. A comparison was made between the Maxwellian distribution curve and the two-temperature distribution curve, where the main population was set to 10~eV and the secondary population was set to 50~eV. Similarly to the previous section (\S\ref{sec:maxwellian}.~Maxwellian Distribution with Suprathermal Beam), the $I$--$V$ curve is plotted in Fig.~\ref{fig:two-temp-IV-curve} along with measurements of the electron temperature using OML~theory. At lower voltages, the tail of the higher-temperature population dominates the electron current, leading to a significant deviation of the electron temperature. As the voltage increases, the curves converge towards the Maxwellian curve. Similarly to the previous section, the small initial change in $p$ shows an appreciable deviation and further increases in $p$ show a lesser effect. \par 

All in all, this illustrates the challenges in the OML~theory in accurately measuring plasma parameters for non-Maxwellian plasmas, especially if not sweeping at sufficiently high voltages to capture all electron populations. Regardless, this diagnostic does not rely on Maxwellian or other assumptions such as local temperature equilibrium (LTE). It is applicable in both ideal (LTE and Maxwellian) and non-ideal (partial-LTE plasmas or non-Maxwellian),  as it only requires that the current–voltage characteristic can be mapped to the local EEDF (electron energy distribution function) under the correct sheath/transport model. Other examples that may affect the validity of the model are strong magnetic fields, complex chemistry (e.g. Fluoride), high-temperature environments (causing electron emission). In this study, these examples were not applicable and did not have to be taken into account for.           

\vspace{-10pt}
%%%%%%%%%%%%%%%%%%%%%%%%%%%%%%%%%%%%%%%%%%%%%%%%%%%%%%%%%%%%%%%%%%%%%%%%%%%%%%%%%%%%
\subsection{mARC~II Data Analysis}
%%%%%%%%%%%%%%%%%%%%%%%%%%%%%%%%%%%%%%%%%%%%%%%%%%%%%%%%%%%%%%%%%%%%%%%%%%%%%%%%%%%%
\vspace{-10pt}

\begin{figure}
\centering
\includegraphics[width=8.5cm]{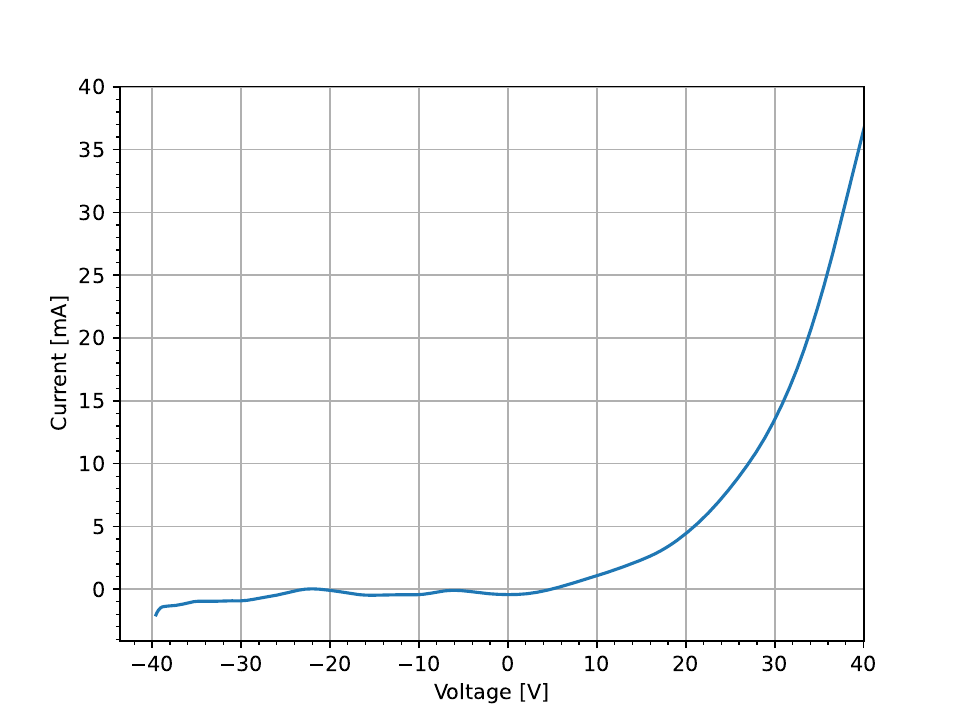}
\caption{An example of a raw characteristic $I$--$V$ curve during insertion at a distance of 50~mm from the nozzle exit plane for Run~029.}
\label{fig:run029-IV-curve}
\end{figure}

The Langmuir probe was implemented in two high-condition runs, and Table~\ref{table:test-matrix} shows the condition parameters for those runs~\cite{Rodrigues2025a, Palmer2025}. Figure~\ref{fig:run029-IV-curve} shows an example of the raw characteristic $I$--$V$ curve obtained from the probe when it is inserted in the plasma flow. There are minor oscillations present within the $I$--$V$ curve. These oscillations are caused by the stray capacitance inherent to the system, as these oscillations appear when there is no plasma discharge. Because the current induced by the stray capacitance is averaged, the smaller oscillations remain after subtracting from the raw data. To remove these oscillations, a Gaussian filter was applied and the result was used as the characteristic $I$--$V$ curve. \par

\begin{table}
\caption{Steady-state test conditions for Runs 029 and 030, for which Langmuir data is presented.}
\begin{ruledtabular}
\begin{tabular}{ccc}
\vspace{1mm}
Conditions&Run 029&Run 030\\
\hline
\vspace{0.5mm}
Set current&200~A&200~A\\
\vspace{0.5mm}
Set mass flow rate (gaseous air)&0.40~g/s&0.45~g/s\\
\vspace{0.5mm}
Test box static pressure&31.2~Pa&35.7~Pa\\
\vspace{0.5mm}
Bulk flow enthalpy&21.0~MJ/kg&22.1~MJ/kg\\\end{tabular}
\end{ruledtabular}
\label{table:test-matrix}
\end{table}

The ion saturation current~$I_i$ was approximated by fitting a line when the probe is at a sufficiently high negative bias. This was then subtracted from the $I$--$V$ curve, providing an approximation of the electron current. An exponential fit was then applied and using Eq.~\eqref{eq:electron-current} the electron temperature was measured\cite{chen2020lecture}: 
\begin{equation}
\label{eq:electron-current}
    I_e = nq_ev_{t}e^{\frac{(V_p - V_s)}{K_BT_e}}~.
\end{equation}

\begin{figure}
\centering
\includegraphics[width=8.5cm]{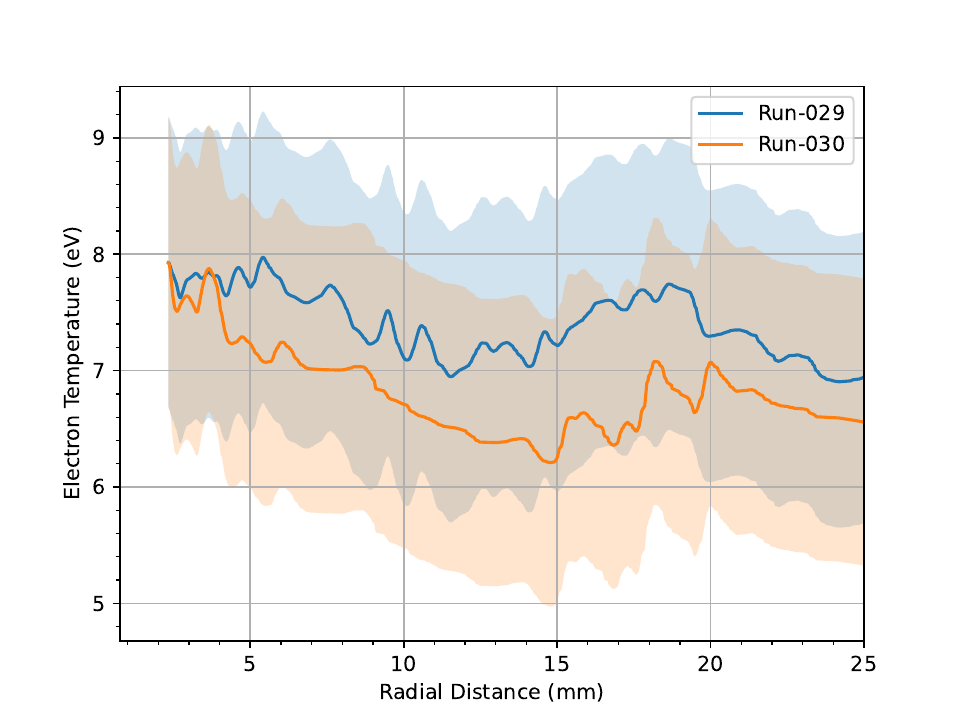}
\caption{Electron temperature radial profile for Run~029~(\solblue) and Run~030~(\solorange) at a distance of 50~mm from the nozzle exit. The shaded region represents the error bounds, determined by calculating the local standard-deviation (1$\sigma$) for every 100 data-points.}
\label{fig:teV}
\end{figure}

Here $n$ is the electron density, $v_t$ is the thermal velocity, $V_p$ is the potential at the probe, and $V_S$ is the space potential. Figure~\ref{fig:teV} illustrates the radial profile of the electron temperature of the flow under steady-state conditions. As one can see, in both runs there is a presence of two peaks in electron temperature, one toward the center of the nozzle and another around 18~mm distance from the center. This may be attributed to the presence of a shock-diamond structure in the flow, leading to a higher number of collisions and thus heating along the plasma jet boundaries. In both runs, the diagnostic measures an electron temperature of approximately 8~eV near the centerline of the nozzle. These values are similar to what was found in previous studies performed at the facility~\cite{mARC-langmuir, johnson1997electron, Diana2021}. Due to restrictions of the test series, the diagnostic could not be validated (e.g., using emission spectroscopy) but this is planned for future efforts. Note that independent validation of the diagnostic was performed against optical emission spectroscopy (OES) measurements conducted in a separate RF-generated plasma experiment. The two methods showed good agreement within measured uncertainty and will be reported in a forthcoming publication. The electron density was calculated by applying the electron temperature measurements and the average ion saturation current $I_i$ to Eq.~\eqref{eq:ion-current}\cite{chen2016introduction}:
\begin{equation}
\label{eq:ion-current}
    I_i = q_enA_p\sqrt{\frac{k_BT_e}{m_i}}e^{-1/2}~.
\end{equation}

\begin{figure}
\centering
\includegraphics[width=8.5cm]{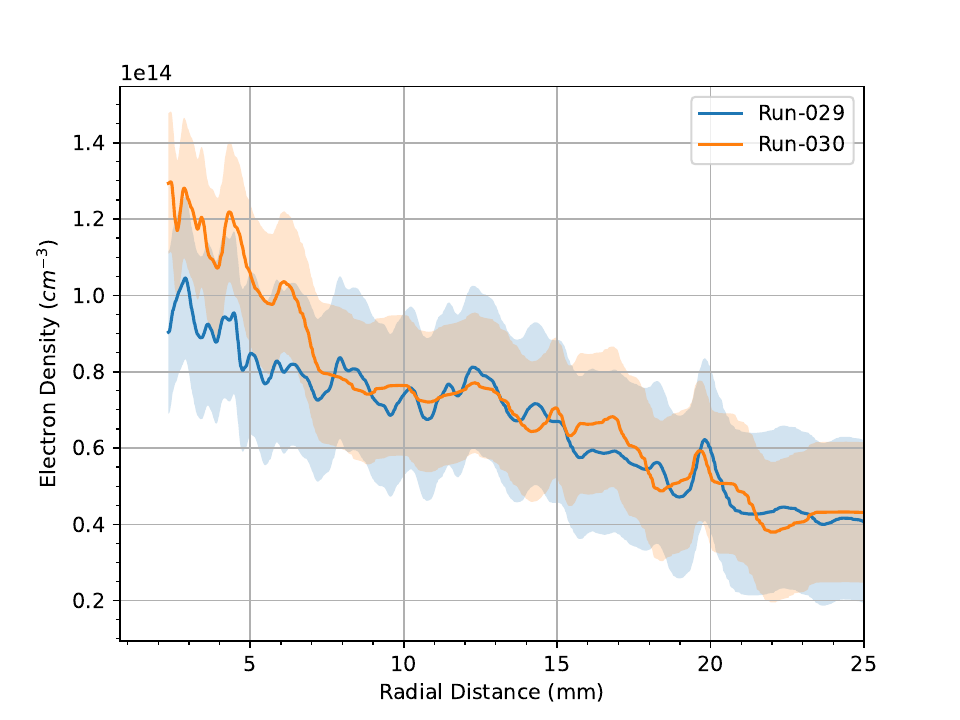}
\caption{Electron density radial profile for Run~029~(\solblue) and Run~030~(\solorange) at a distance of 50~mm from the nozzle exit. The shaded region represents the error bounds, determined by calculating the local standard-deviation (1$\sigma$) for every 100 data-points.}
\label{fig:ne}
\end{figure}

Here is $m_i$ is the mass of the ion. Figure~\ref{fig:ne} illustrates the radial profile of the electron density under steady-state conditions. It showcases a semi-linear increase in electron density towards the centerline of the nozzle flow. The error associated with the figures was determined using the standard deviation calculated on localized segments of 100 data points. This approach ensures that variability within smaller regions of the data set is accurately captured. However, there is still uncertainty in the absolute error associated with these measurements, and there will be future efforts to quantify the error via error propagation.

The fast-sweeping characteristic of the probe provides the capability to measure higher-frequency fluctuations in these plasma parameters, which, in comparison, is not possible for traditional sweeping probes~\cite{400khz, Prevosto_SSLP}. The frequency spectrum of the electron temperature was determined and showcased peaks in $\sim12.1$~kHz and $\sim20.7$~kHz. This is aligned with other studies have found in similar facilities due to the oscillations caused by the arcing~\cite{thomas1997fluctuation}. There will be efforts to investigate if these oscillations are real or produced by noise. Note that since the electron density is calculated using the measured electron temperature, the same oscillatory peaks are present in the electron density measurements.

\begin{figure}
\centering
\includegraphics[width=8.5cm]{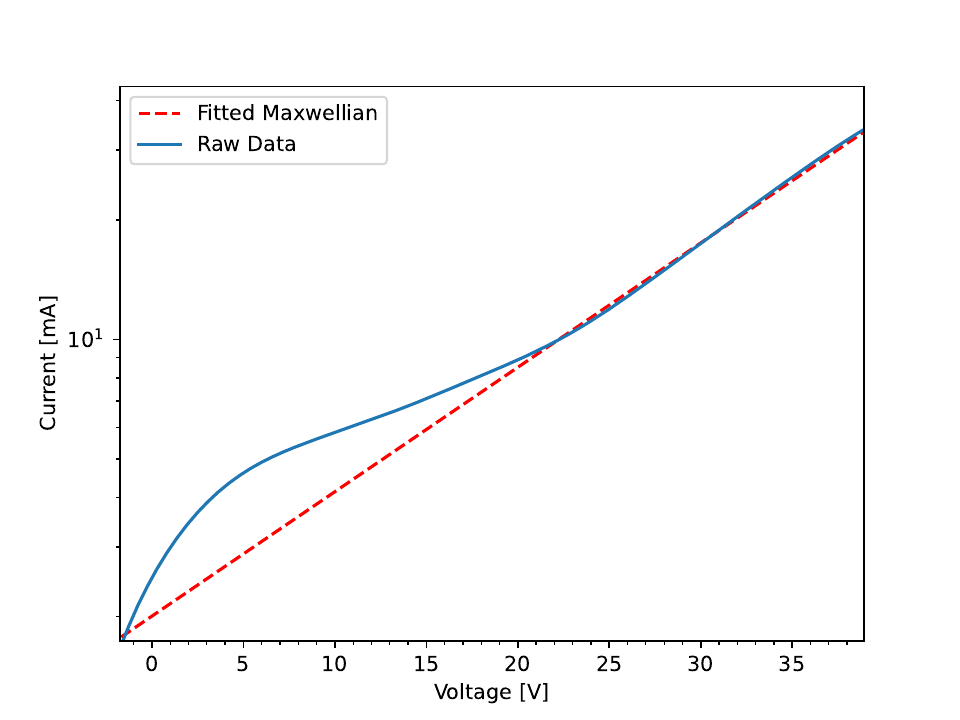}
\caption{An example of the raw electron current at logarithmic scale. In an ideal Maxwellian plasma, the electron current would follow a linear profile (\dashred). However, in majority of sweeps there is a presence of a bump occurring at the lower voltages. This possibly indicates a non-Maxwellian plasma, specifically the presence of a suprathermal beam that is slower than the primary population of electrons.}
\label{fig:suprathermal_data}
\end{figure}

Additionally, in majority of the sweeps, there is a bump is identified at the lower voltages when the logarithmic of the electron current data is taken. An example is shown in Figure~\ref{fig:suprathermal_data}. This looks similar to the bumps illustrated in Figure~\ref{fig:maxwellian-IV-curve} for a Maxwellian $+$ suprathermal beam and thus possibly indicates the presence of a subthermal beam of electrons within the plasma flow. The peak of these bumps occur around 5~V which would indicate an average velocity of $1.33\times10^6$~m/s. Due to this bump occurring at lower voltages, this would indicate that it would theoretically be slower than the primary population of electrons. A possible explanation for this “drifting bump” is that when a plasma jet forms a positive potential core relative to the vacuum edge, electrons attempting to escape may be reflected back \textit{if} their kinetic energy is below the potential barrier~\cite{HairapetianStenzel1988}. These reflected electrons accumulate and produce a distinct low-energy bump in the velocity distribution, often drifting inward relative to the bulk expansion. Currently, there are efforts to better characterize and understand these bumps, including whether they are a physical measurement or an artifact of noise -- however, this is outside of scope for this paper.

\vspace{-10pt}

\vspace{-10pt}

\vspace{10pt}
%%%%%%%%%%%%%%%%%%%%%%%%%%%%%%%%%%%%%%%%%%%%%%%%%%%%%%%%%%%%%%%%%%%%%%%%%%%%%%%%%%%%
\section{Conclusions and Future Work}
%%%%%%%%%%%%%%%%%%%%%%%%%%%%%%%%%%%%%%%%%%%%%%%%%%%%%%%%%%%%%%%%%%%%%%%%%%%%%%%%%%%%
\vspace{-10pt}

In response to the primary objective of this project, the fast-sweeping Langmuir probe system provides measurements of the plasma's characteristic $I$--$V$ curves in the extreme aerothermal conditions found in arc jet facilities. In one test, the diagnostic provided electron temperature and density radial profiles, showing a maximum of 8~eV and $1.3$~$\times$~$10^{14}$~cm$^{-3}$, respectively. Due to its higher temporal resolution, this diagnostics opens an avenue to measuring the evolution of these plasma parameters and overall better understand uncertainty in these extreme aero-thermal conditions. This cost-effective (estimated cost of the components and the printed circuit board manufacturing is 184~USD), open-source diagnostic opens a door to new capabilities for similar ground-testing facilities and new insight to the plasma's spatio-temporal behaviors. \par 

This is considered to be the first design iteration, and there have been ongoing efforts to improve the design. This includes implementing capacitive compensation, such as a dummy probe, to minimize the current induced by stray capacitance when sweeping at higher frequencies. Additionally, the technique used to measure the voltage across the shunt resistor will be improved by implementing a differential amplifier or similar method, increasing the accuracy of the current measurements. Furthermore, there is motivation to redesign sub-sections on the circuit design such that higher sweeping frequencies can be obtained for testing facilities with short-period plasma discharges (lasting less than millisecond timescales). \par  

Lastly, there are plans to continue the efforts to fully characterize the plasma produced at the mARC~II facility. This includes efforts to analyze the fluctuations found within these parameters because of their importance in understanding energy transport and dissipation. Langmuir diagnostics will provide insight not only into fundamental plasma parameters but also into the presence of non-Maxwellian distributions. To achieve this, we plan to numerically simulate small-scale behaviors and develop a robust model for non-Maxwellian distributions for low-ionization, collisional plasmas. \par 

\vspace{-10pt}
%%%%%%%%%%%%%%%%%%%%%%%%%%%%%%%%%%%%%%%%%%%%%%%%%%%%%%%%%%%%%%%%%%%%%%%%%%%%%%%%%%%%
\begin{acknowledgments}
%%%%%%%%%%%%%%%%%%%%%%%%%%%%%%%%%%%%%%%%%%%%%%%%%%%%%%%%%%%%%%%%%%%%%%%%%%%%%%%%%%%%
Jocelino Rodrigues is the recipient of a NASA Postdoctoral Program~(NPP) research fellowship at NASA Ames Research Center, administered by Oak Ridge Associated Universities~(ORAU) under the NASA contract 80HQTR21CA005. The authors are grateful to the mARC~II facility team (Ramon Martinez, Ryan Chung, Daniel Philippidis) for their support in implementing this new diagnostic. In particular, the authors appreciate Joe Hartman and Dr.~Megan MacDonald for their insightful guidance in understanding arc jet behaviors. 
\end{acknowledgments}

\vspace{-10pt}
%%%%%%%%%%%%%%%%%%%%%%%%%%%%%%%%%%%%%%%%%%%%%%%%%%%%%%%%%%%%%%%%%%%%%%%%%%%%%%%%%%%%
\section*{Data Availability Statement}
%%%%%%%%%%%%%%%%%%%%%%%%%%%%%%%%%%%%%%%%%%%%%%%%%%%%%%%%%%%%%%%%%%%%%%%%%%%%%%%%%%%%
\vspace{-10pt}

In line with one of the primary motivations of this paper, the design of the Langmuir circuit is openly available on the LEGOLAS GitHub repository: \url{https://github.com/Cool-Whiskers/LEGOLAS}. The raw data were generated at the mARC~II facility. Data derived to support the findings of this study are available from the corresponding author upon reasonable request.

\vspace{-10pt}
%%%%%%%%%%%%%%%%%%%%%%%%%%%%%%%%%%%%%%%%%%%%%%%%%%%%%%%%%%%%%%%%%%%%%%%%%%%%%%%%%%%%
\section*{References}
%%%%%%%%%%%%%%%%%%%%%%%%%%%%%%%%%%%%%%%%%%%%%%%%%%%%%%%%%%%%%%%%%%%%%%%%%%%%%%%%%%%%
\vspace{-10pt}

\nocite{*}
\bibliography{aipsamp}% Produces the bibliography via BibTeX.

\providecommand{\noopsort}[1]{}\providecommand{\singleletter}[1]{#1}%
\begin{thebibliography}{45}%
\makeatletter
\providecommand \@ifxundefined [1]{%
 \@ifx{#1\undefined}
}%
\providecommand \@ifnum [1]{%
 \ifnum #1\expandafter \@firstoftwo
 \else \expandafter \@secondoftwo
 \fi
}%
\providecommand \@ifx [1]{%
 \ifx #1\expandafter \@firstoftwo
 \else \expandafter \@secondoftwo
 \fi
}%
\providecommand \natexlab [1]{#1}%
\providecommand \enquote  [1]{``#1''}%
\providecommand \bibnamefont  [1]{#1}%
\providecommand \bibfnamefont [1]{#1}%
\providecommand \citenamefont [1]{#1}%
\providecommand \href@noop [0]{\@secondoftwo}%
\providecommand \href [0]{\begingroup \@sanitize@url \@href}%
\providecommand \@href[1]{\@@startlink{#1}\@@href}%
\providecommand \@@href[1]{\endgroup#1\@@endlink}%
\providecommand \@sanitize@url [0]{\catcode `\\12\catcode `\$12\catcode `\&12\catcode `\#12\catcode `\^12\catcode `\_12\catcode `\%12\relax}%
\providecommand \@@startlink[1]{}%
\providecommand \@@endlink[0]{}%
\providecommand \url  [0]{\begingroup\@sanitize@url \@url }%
\providecommand \@url [1]{\endgroup\@href {#1}{\urlprefix }}%
\providecommand \urlprefix  [0]{URL }%
\providecommand \Eprint [0]{\href }%
\providecommand \doibase [0]{http://dx.doi.org/}%
\providecommand \selectlanguage [0]{\@gobble}%
\providecommand \bibinfo  [0]{\@secondoftwo}%
\providecommand \bibfield  [0]{\@secondoftwo}%
\providecommand \translation [1]{[#1]}%
\providecommand \BibitemOpen [0]{}%
\providecommand \bibitemStop [0]{}%
\providecommand \bibitemNoStop [0]{.\EOS\space}%
\providecommand \EOS [0]{\spacefactor3000\relax}%
\providecommand \BibitemShut  [1]{\csname bibitem#1\endcsname}%
\let\auto@bib@innerbib\@empty
%</preamble>
\bibitem [{\citenamefont {Huddlestone}\ and\ \citenamefont {Leonard}(1965)}]{huddlestone1965langmuir}%
  \BibitemOpen
  \bibfield  {author} {\bibinfo {author} {\bibfnamefont {R.~H.}\ \bibnamefont {Huddlestone}}\ and\ \bibinfo {author} {\bibfnamefont {S.~L.}\ \bibnamefont {Leonard}},\ }\bibfield  {title} {\enquote {\bibinfo {title} {Langmuir probes},}\ }in\ \href {https://doi.org/10.1016/B978-1-4832-2897-8.50010-0} {\emph {\bibinfo {booktitle} {Plasma Diagnostic Techniques}}},\ \bibinfo {editor} {edited by\ \bibinfo {editor} {\bibfnamefont {R.~H.}\ \bibnamefont {Huddlestone}}\ and\ \bibinfo {editor} {\bibfnamefont {S.~L.}\ \bibnamefont {Leonard}}}\ (\bibinfo  {publisher} {Academic Press},\ \bibinfo {year} {1965})\ pp.\ \bibinfo {pages} {113--200}\BibitemShut {NoStop}%
\bibitem [{\citenamefont {Chen}(2016)}]{chen2016introduction}%
  \BibitemOpen
  \bibfield  {author} {\bibinfo {author} {\bibfnamefont {F.~F.}\ \bibnamefont {Chen}},\ }\href {\doibase 10.1007/978-3-319-22309-4} {\emph {\bibinfo {title} {Introduction to Plasma Physics and Controlled Fusion}}},\ \bibinfo {edition} {3rd}\ ed.\ (\bibinfo  {publisher} {Springer},\ \bibinfo {year} {2016})\BibitemShut {NoStop}%
\bibitem [{\citenamefont {Codron}\ and\ \citenamefont {Nawaz}(2013)}]{mARC-langmuir}%
  \BibitemOpen
  \bibfield  {author} {\bibinfo {author} {\bibfnamefont {D.~A.}\ \bibnamefont {Codron}}\ and\ \bibinfo {author} {\bibfnamefont {A.}~\bibnamefont {Nawaz}},\ }\bibfield  {title} {\enquote {\bibinfo {title} {Radial profiles of the plasma electron characteristics in a 30kw arc jet},}\ }\href@noop {} {\bibfield  {journal} {\bibinfo  {journal} {44th AIAA Plasmadynamics and Lasers Conference}\ } (\bibinfo {year} {2013})},\ \bibinfo {note} {{DOI:10.2514/6.2013-3128}}\BibitemShut {NoStop}%
\bibitem [{\citenamefont {Mart{\'i}nez}\ \emph {et~al.}(2022)\citenamefont {Mart{\'i}nez}, \citenamefont {Rodr{\'i}guez}, \citenamefont {Pedr{\'o}s}, \citenamefont {Ballesteros},\ and\ \citenamefont {Mart{\'i}nez}}]{emmissive}%
  \BibitemOpen
  \bibfield  {author} {\bibinfo {author} {\bibfnamefont {R.}~\bibnamefont {Mart{\'i}nez}}, \bibinfo {author} {\bibfnamefont {J.}~\bibnamefont {Rodr{\'i}guez}}, \bibinfo {author} {\bibfnamefont {M.~A.}\ \bibnamefont {Pedr{\'o}s}}, \bibinfo {author} {\bibfnamefont {J.}~\bibnamefont {Ballesteros}}, \ and\ \bibinfo {author} {\bibfnamefont {J.~M.}\ \bibnamefont {Mart{\'i}nez}},\ }\bibfield  {title} {\enquote {\bibinfo {title} {Floating potential method using a thermionic emissive probe including an ionizing and collisional presheath},}\ }\href {\doibase 10.1088/1361-6595/ac8b0e} {\bibfield  {journal} {\bibinfo  {journal} {Plasma Sources Science and Technology}\ }\textbf {\bibinfo {volume} {31}},\ \bibinfo {pages} {095012} (\bibinfo {year} {2022})}\BibitemShut {NoStop}%
\bibitem [{\citenamefont {Andrenucci}, \citenamefont {Martín-Saravia},\ and\ \citenamefont {Ahedo}(2019)}]{hall-thruster}%
  \BibitemOpen
  \bibfield  {author} {\bibinfo {author} {\bibfnamefont {M.}~\bibnamefont {Andrenucci}}, \bibinfo {author} {\bibfnamefont {M.}~\bibnamefont {Martín-Saravia}}, \ and\ \bibinfo {author} {\bibfnamefont {E.}~\bibnamefont {Ahedo}},\ }\bibfield  {title} {\enquote {\bibinfo {title} {Plasma characterization in hall thrusters by langmuir probes},}\ }\href {\doibase 10.1088/1748-0221/14/05/C05011} {\bibfield  {journal} {\bibinfo  {journal} {Journal of Instrumentation}\ }\textbf {\bibinfo {volume} {14}},\ \bibinfo {pages} {C05011} (\bibinfo {year} {2019})}\BibitemShut {NoStop}%
\bibitem [{\citenamefont {Haw}, \citenamefont {Seo},\ and\ \citenamefont {Bellan}(2019)}]{haw2019laboratory}%
  \BibitemOpen
  \bibfield  {author} {\bibinfo {author} {\bibfnamefont {M.~A.}\ \bibnamefont {Haw}}, \bibinfo {author} {\bibfnamefont {B.}~\bibnamefont {Seo}}, \ and\ \bibinfo {author} {\bibfnamefont {P.~M.}\ \bibnamefont {Bellan}},\ }\bibfield  {title} {\enquote {\bibinfo {title} {Laboratory measurement of large-amplitude whistler pulses generated by fast magnetic reconnection},}\ }\href {\doibase 10.1029/2019GL082621} {\bibfield  {journal} {\bibinfo  {journal} {Geophysical Research Letters}\ }\textbf {\bibinfo {volume} {46}},\ \bibinfo {pages} {7342--7351} (\bibinfo {year} {2019})}\BibitemShut {NoStop}%
\bibitem [{\citenamefont {Efthimion}\ \emph {et~al.}(2005)\citenamefont {Efthimion}, \citenamefont {Gilson}, \citenamefont {Grisham}, \citenamefont {Davidson}, \citenamefont {Yu}, \citenamefont {Waldron},\ and\ \citenamefont {Logan}}]{efthimion2005development}%
  \BibitemOpen
  \bibfield  {author} {\bibinfo {author} {\bibfnamefont {P.~C.}\ \bibnamefont {Efthimion}}, \bibinfo {author} {\bibfnamefont {E.~P.}\ \bibnamefont {Gilson}}, \bibinfo {author} {\bibfnamefont {L.}~\bibnamefont {Grisham}}, \bibinfo {author} {\bibfnamefont {R.~C.}\ \bibnamefont {Davidson}}, \bibinfo {author} {\bibfnamefont {S.}~\bibnamefont {Yu}}, \bibinfo {author} {\bibfnamefont {W.}~\bibnamefont {Waldron}}, \ and\ \bibinfo {author} {\bibfnamefont {B.~G.}\ \bibnamefont {Logan}},\ }\href {https://escholarship.org/content/qt39p3z04h/qt39p3z04h.pdf} {\enquote {\bibinfo {title} {Development of a one-meter plasma source for heavy ion beam charge neutralization},}\ }\bibinfo {type} {Tech. Rep.}\ (\bibinfo  {institution} {Lawrence Berkeley National Laboratory},\ \bibinfo {year} {2005})\BibitemShut {NoStop}%
\bibitem [{\citenamefont {Piejak}, \citenamefont {Godyak},\ and\ \citenamefont {Alexandrovich}(2016)}]{fusion}%
  \BibitemOpen
  \bibfield  {author} {\bibinfo {author} {\bibfnamefont {R.~B.}\ \bibnamefont {Piejak}}, \bibinfo {author} {\bibfnamefont {V.~A.}\ \bibnamefont {Godyak}}, \ and\ \bibinfo {author} {\bibfnamefont {B.~M.}\ \bibnamefont {Alexandrovich}},\ }\bibfield  {title} {\enquote {\bibinfo {title} {Advances in langmuir probe diagnostics of the plasma potential and electron-energy distribution function in magnetized plasma},}\ }\href {\doibase 10.1088/0963-0252/25/3/033001} {\bibfield  {journal} {\bibinfo  {journal} {Plasma Sources Science and Technology}\ }\textbf {\bibinfo {volume} {25}},\ \bibinfo {pages} {033001} (\bibinfo {year} {2016})}\BibitemShut {NoStop}%
\bibitem [{\citenamefont {Katz}\ \emph {et~al.}(2008)\citenamefont {Katz}, \citenamefont {Egedal}, \citenamefont {Fox},\ and\ \citenamefont {Porkolab}}]{katz2008experiments}%
  \BibitemOpen
  \bibfield  {author} {\bibinfo {author} {\bibfnamefont {N.}~\bibnamefont {Katz}}, \bibinfo {author} {\bibfnamefont {J.}~\bibnamefont {Egedal}}, \bibinfo {author} {\bibfnamefont {W.}~\bibnamefont {Fox}}, \ and\ \bibinfo {author} {\bibfnamefont {M.}~\bibnamefont {Porkolab}},\ }\bibfield  {title} {\enquote {\bibinfo {title} {Experiments on the propagation of plasma filaments},}\ }\href {\doibase 10.1103/PhysRevLett.101.015003} {\bibfield  {journal} {\bibinfo  {journal} {Physical Review Letters}\ }\textbf {\bibinfo {volume} {101}},\ \bibinfo {pages} {015003} (\bibinfo {year} {2008})}\BibitemShut {NoStop}%
\bibitem [{\citenamefont {Berg}\ \emph {et~al.}(2019)\citenamefont {Berg}, \citenamefont {Wahlund}, \citenamefont {Andr{\'e}}, \citenamefont {Eriksson}, \citenamefont {Bekkeng}, \citenamefont {Dehmel}, \citenamefont {Norberg}, \citenamefont {Eriksen}, \citenamefont {Trondsen}, \citenamefont {Gr{\o}tte}, \citenamefont {Holsb{\o}}, \citenamefont {Lindem}, \citenamefont {L{\o}vhaug}, \citenamefont {Lindem}, \citenamefont {Jacobsen}, \citenamefont {H{\o}nsi}, \citenamefont {Lybekk},\ and\ \citenamefont {Moen}}]{spacecraft}%
  \BibitemOpen
  \bibfield  {author} {\bibinfo {author} {\bibfnamefont {H.}~\bibnamefont {Berg}}, \bibinfo {author} {\bibfnamefont {J.~E.}\ \bibnamefont {Wahlund}}, \bibinfo {author} {\bibfnamefont {M.}~\bibnamefont {Andr{\'e}}}, \bibinfo {author} {\bibfnamefont {A.}~\bibnamefont {Eriksson}}, \bibinfo {author} {\bibfnamefont {J.~K.}\ \bibnamefont {Bekkeng}}, \bibinfo {author} {\bibfnamefont {L.-E.}\ \bibnamefont {Dehmel}}, \bibinfo {author} {\bibfnamefont {O.~P.}\ \bibnamefont {Norberg}}, \bibinfo {author} {\bibfnamefont {T.}~\bibnamefont {Eriksen}}, \bibinfo {author} {\bibfnamefont {E.}~\bibnamefont {Trondsen}}, \bibinfo {author} {\bibfnamefont {T.-A.}\ \bibnamefont {Gr{\o}tte}}, \bibinfo {author} {\bibfnamefont {E.}~\bibnamefont {Holsb{\o}}}, \bibinfo {author} {\bibfnamefont {T.}~\bibnamefont {Lindem}}, \bibinfo {author} {\bibfnamefont {E.~S.}\ \bibnamefont {L{\o}vhaug}}, \bibinfo {author} {\bibfnamefont {T.~M.}\ \bibnamefont {Lindem}}, \bibinfo {author} {\bibfnamefont {K.~S.}\ \bibnamefont {Jacobsen}}, \bibinfo {author}
  {\bibfnamefont {D.~T.}\ \bibnamefont {H{\o}nsi}}, \bibinfo {author} {\bibfnamefont {B.}~\bibnamefont {Lybekk}}, \ and\ \bibinfo {author} {\bibfnamefont {J.~I.}\ \bibnamefont {Moen}},\ }\bibfield  {title} {\enquote {\bibinfo {title} {Multi-needle langmuir probe system for electron density measurements and active spacecraft potential control on cubesats},}\ }\href {\doibase 10.1109/TPS.2019.2894466} {\bibfield  {journal} {\bibinfo  {journal} {IEEE Transactions on Plasma Science}\ }\textbf {\bibinfo {volume} {47}},\ \bibinfo {pages} {1512--1531} (\bibinfo {year} {2019})}\BibitemShut {NoStop}%
\bibitem [{\citenamefont {Irimiciuc}\ \emph {et~al.}(2021)\citenamefont {Irimiciuc}, \citenamefont {Chertopalov}, \citenamefont {Lancok},\ and\ \citenamefont {Craciun}}]{semiconductor}%
  \BibitemOpen
  \bibfield  {author} {\bibinfo {author} {\bibfnamefont {S.~A.}\ \bibnamefont {Irimiciuc}}, \bibinfo {author} {\bibfnamefont {S.}~\bibnamefont {Chertopalov}}, \bibinfo {author} {\bibfnamefont {J.}~\bibnamefont {Lancok}}, \ and\ \bibinfo {author} {\bibfnamefont {V.}~\bibnamefont {Craciun}},\ }\bibfield  {title} {\enquote {\bibinfo {title} {Langmuir probe technique for plasma characterization during pulsed laser deposition process},}\ }\href {\doibase 10.3390/coatings11070762} {\bibfield  {journal} {\bibinfo  {journal} {Coatings}\ }\textbf {\bibinfo {volume} {11}},\ \bibinfo {pages} {762} (\bibinfo {year} {2021})}\BibitemShut {NoStop}%
\bibitem [{\citenamefont {Polzin}\ \emph {et~al.}(2019)\citenamefont {Polzin}, \citenamefont {Blumhagen}, \citenamefont {Sherrod},\ and\ \citenamefont {Moeller}}]{nasa-triple}%
  \BibitemOpen
  \bibfield  {author} {\bibinfo {author} {\bibfnamefont {K.~A.}\ \bibnamefont {Polzin}}, \bibinfo {author} {\bibfnamefont {E.}~\bibnamefont {Blumhagen}}, \bibinfo {author} {\bibfnamefont {A.~C.}\ \bibnamefont {Sherrod}}, \ and\ \bibinfo {author} {\bibfnamefont {T.~M.}\ \bibnamefont {Moeller}},\ }\bibfield  {title} {\enquote {\bibinfo {title} {Behavior of triple langmuir probes in non-equilibrium plasmas},}\ }in\ \href {https://ntrs.nasa.gov/api/citations/20190030439/downloads/20190030439.pdf} {\emph {\bibinfo {booktitle} {AIAA Propulsion and Energy Forum}}}\ (\bibinfo  {publisher} {American Institute of Aeronautics and Astronautics},\ \bibinfo {address} {Indianapolis, IN},\ \bibinfo {year} {2019})\ \bibinfo {note} {nASA Technical Report 20190030439}\BibitemShut {NoStop}%
\bibitem [{\citenamefont {Schubert}\ \emph {et~al.}(2007)\citenamefont {Schubert}, \citenamefont {Endler}, \citenamefont {Thomsen},\ and\ \citenamefont {Team}}]{fast-swept-spatiotemporal-array}%
  \BibitemOpen
  \bibfield  {author} {\bibinfo {author} {\bibfnamefont {M.}~\bibnamefont {Schubert}}, \bibinfo {author} {\bibfnamefont {M.}~\bibnamefont {Endler}}, \bibinfo {author} {\bibfnamefont {H.}~\bibnamefont {Thomsen}}, \ and\ \bibinfo {author} {\bibfnamefont {W.-A.}\ \bibnamefont {Team}},\ }\bibfield  {title} {\enquote {\bibinfo {title} {Spatiotemporal temperature fluctuation measurements by means of a fast swept langmuir probe array},}\ }\href {\doibase 10.1063/1.2740785} {\bibfield  {journal} {\bibinfo  {journal} {Review of Scientific Instruments}\ }\textbf {\bibinfo {volume} {78}},\ \bibinfo {pages} {053505} (\bibinfo {year} {2007})}\BibitemShut {NoStop}%
\bibitem [{\citenamefont {Troll}\ \emph {et~al.}(2010)\citenamefont {Troll}, \citenamefont {Conde}, \citenamefont {Criado}, \citenamefont {Donoso},\ and\ \citenamefont {Herdrich}}]{langmuir-low-ionization}%
  \BibitemOpen
  \bibfield  {author} {\bibinfo {author} {\bibfnamefont {O.}~\bibnamefont {Troll}}, \bibinfo {author} {\bibfnamefont {L.}~\bibnamefont {Conde}}, \bibinfo {author} {\bibfnamefont {E.}~\bibnamefont {Criado}}, \bibinfo {author} {\bibfnamefont {J.~M.}\ \bibnamefont {Donoso}}, \ and\ \bibinfo {author} {\bibfnamefont {G.}~\bibnamefont {Herdrich}},\ }\bibfield  {title} {\enquote {\bibinfo {title} {Measurements of plasma properties using fast sweep langmuir probes in unmagnetized weakly ionized plasmas},}\ }\href {\doibase 10.1002/ctpp.201010138} {\bibfield  {journal} {\bibinfo  {journal} {Contributions to Plasma Physics}\ }\textbf {\bibinfo {volume} {50}},\ \bibinfo {pages} {819--823} (\bibinfo {year} {2010})}\BibitemShut {NoStop}%
\bibitem [{\citenamefont {Nold}\ \emph {et~al.}(2012)\citenamefont {Nold}, \citenamefont {Ribeiro}, \citenamefont {Ramisch}, \citenamefont {Huang}, \citenamefont {M{\"u}ller}, \citenamefont {Scott}, \citenamefont {Stroth},\ and\ \citenamefont {Team}}]{turbulence}%
  \BibitemOpen
  \bibfield  {author} {\bibinfo {author} {\bibfnamefont {B.}~\bibnamefont {Nold}}, \bibinfo {author} {\bibfnamefont {T.~T.}\ \bibnamefont {Ribeiro}}, \bibinfo {author} {\bibfnamefont {M.}~\bibnamefont {Ramisch}}, \bibinfo {author} {\bibfnamefont {Z.}~\bibnamefont {Huang}}, \bibinfo {author} {\bibfnamefont {H.~W.}\ \bibnamefont {M{\"u}ller}}, \bibinfo {author} {\bibfnamefont {B.~D.}\ \bibnamefont {Scott}}, \bibinfo {author} {\bibfnamefont {U.}~\bibnamefont {Stroth}}, \ and\ \bibinfo {author} {\bibfnamefont {A.~U.}\ \bibnamefont {Team}},\ }\bibfield  {title} {\enquote {\bibinfo {title} {Influence of temperature fluctuations on plasma turbulence investigations with langmuir probes},}\ }\href {\doibase 10.1088/1367-2630/14/6/063022} {\bibfield  {journal} {\bibinfo  {journal} {New Journal of Physics}\ }\textbf {\bibinfo {volume} {14}},\ \bibinfo {pages} {063022} (\bibinfo {year} {2012})}\BibitemShut {NoStop}%
\bibitem [{\citenamefont {Hu}\ \emph {et~al.}(2021)\citenamefont {Hu}, \citenamefont {Yoo}, \citenamefont {Ji}, \citenamefont {Goodman},\ and\ \citenamefont {Wu}}]{hu2021probe}%
  \BibitemOpen
  \bibfield  {author} {\bibinfo {author} {\bibfnamefont {Y.}~\bibnamefont {Hu}}, \bibinfo {author} {\bibfnamefont {J.}~\bibnamefont {Yoo}}, \bibinfo {author} {\bibfnamefont {H.}~\bibnamefont {Ji}}, \bibinfo {author} {\bibfnamefont {A.}~\bibnamefont {Goodman}}, \ and\ \bibinfo {author} {\bibfnamefont {X.}~\bibnamefont {Wu}},\ }\bibfield  {title} {\enquote {\bibinfo {title} {Probe measurements of electric field and electron density fluctuations at megahertz frequencies using in-shaft miniature circuits},}\ }\href {\doibase 10.1063/5.0035135} {\bibfield  {journal} {\bibinfo  {journal} {Review of Scientific Instruments}\ }\textbf {\bibinfo {volume} {92}},\ \bibinfo {pages} {033534} (\bibinfo {year} {2021})}\BibitemShut {NoStop}%
\bibitem [{\citenamefont {Bose}\ \emph {et~al.}(2018)\citenamefont {Bose}, \citenamefont {Singh}, \citenamefont {Upadhyay},\ and\ \citenamefont {Shyam}}]{pulsed-plasma}%
  \BibitemOpen
  \bibfield  {author} {\bibinfo {author} {\bibfnamefont {A.}~\bibnamefont {Bose}}, \bibinfo {author} {\bibfnamefont {S.~K.}\ \bibnamefont {Singh}}, \bibinfo {author} {\bibfnamefont {A.~K.}\ \bibnamefont {Upadhyay}}, \ and\ \bibinfo {author} {\bibfnamefont {A.}~\bibnamefont {Shyam}},\ }\bibfield  {title} {\enquote {\bibinfo {title} {Study of plasma parameters in a pulsed plasma accelerator using triple langmuir probe},}\ }\href {\doibase 10.1063/1.5017845} {\bibfield  {journal} {\bibinfo  {journal} {Physics of Plasmas}\ }\textbf {\bibinfo {volume} {25}},\ \bibinfo {pages} {013532} (\bibinfo {year} {2018})}\BibitemShut {NoStop}%
\bibitem [{\citenamefont {Gao}\ \emph {et~al.}(2022)\citenamefont {Gao}, \citenamefont {Kraus}, \citenamefont {Hill}, \citenamefont {Schneider}, \citenamefont {Christopherson}, \citenamefont {Bachmann}, \citenamefont {Bitter}, \citenamefont {Efthimion}, \citenamefont {Pablant}, \citenamefont {Betti}, \citenamefont {Thomas}, \citenamefont {Thorn}, \citenamefont {MacPhee}, \citenamefont {Khan}, \citenamefont {Kauffman}, \citenamefont {Liedahl}, \citenamefont {Chen}, \citenamefont {Bradley}, \citenamefont {Kilkenny}, \citenamefont {Lahmann}, \citenamefont {Stambulchik},\ and\ \citenamefont {Maron}}]{gao2022hot}%
  \BibitemOpen
  \bibfield  {author} {\bibinfo {author} {\bibfnamefont {L.}~\bibnamefont {Gao}}, \bibinfo {author} {\bibfnamefont {B.~F.}\ \bibnamefont {Kraus}}, \bibinfo {author} {\bibfnamefont {K.~W.}\ \bibnamefont {Hill}}, \bibinfo {author} {\bibfnamefont {M.~B.}\ \bibnamefont {Schneider}}, \bibinfo {author} {\bibfnamefont {A.}~\bibnamefont {Christopherson}}, \bibinfo {author} {\bibfnamefont {B.}~\bibnamefont {Bachmann}}, \bibinfo {author} {\bibfnamefont {M.}~\bibnamefont {Bitter}}, \bibinfo {author} {\bibfnamefont {P.}~\bibnamefont {Efthimion}}, \bibinfo {author} {\bibfnamefont {N.}~\bibnamefont {Pablant}}, \bibinfo {author} {\bibfnamefont {R.}~\bibnamefont {Betti}}, \bibinfo {author} {\bibfnamefont {C.}~\bibnamefont {Thomas}}, \bibinfo {author} {\bibfnamefont {D.}~\bibnamefont {Thorn}}, \bibinfo {author} {\bibfnamefont {A.~G.}\ \bibnamefont {MacPhee}}, \bibinfo {author} {\bibfnamefont {S.}~\bibnamefont {Khan}}, \bibinfo {author} {\bibfnamefont {R.}~\bibnamefont {Kauffman}}, \bibinfo {author} {\bibfnamefont
  {D.}~\bibnamefont {Liedahl}}, \bibinfo {author} {\bibfnamefont {H.}~\bibnamefont {Chen}}, \bibinfo {author} {\bibfnamefont {D.}~\bibnamefont {Bradley}}, \bibinfo {author} {\bibfnamefont {J.}~\bibnamefont {Kilkenny}}, \bibinfo {author} {\bibfnamefont {B.}~\bibnamefont {Lahmann}}, \bibinfo {author} {\bibfnamefont {E.}~\bibnamefont {Stambulchik}}, \ and\ \bibinfo {author} {\bibfnamefont {Y.}~\bibnamefont {Maron}},\ }\bibfield  {title} {\enquote {\bibinfo {title} {Hot spot evolution measured by high-resolution x-ray spectroscopy at the national ignition facility},}\ }\href {\doibase 10.1103/PhysRevLett.128.185002} {\bibfield  {journal} {\bibinfo  {journal} {Physical Review Letters}\ }\textbf {\bibinfo {volume} {128}},\ \bibinfo {pages} {185002} (\bibinfo {year} {2022})}\BibitemShut {NoStop}%
\bibitem [{\citenamefont {Lobbia}\ and\ \citenamefont {Gallimore}(2010)}]{temporal-limits}%
  \BibitemOpen
  \bibfield  {author} {\bibinfo {author} {\bibfnamefont {R.~B.}\ \bibnamefont {Lobbia}}\ and\ \bibinfo {author} {\bibfnamefont {A.~D.}\ \bibnamefont {Gallimore}},\ }\bibfield  {title} {\enquote {\bibinfo {title} {Temporal limits of a rapidly swept langmuir probe},}\ }\href {\doibase 10.1063/1.3449588} {\bibfield  {journal} {\bibinfo  {journal} {Physics of Plasmas}\ }\textbf {\bibinfo {volume} {17}},\ \bibinfo {pages} {073502} (\bibinfo {year} {2010})}\BibitemShut {NoStop}%
\bibitem [{\citenamefont {Chiodini}, \citenamefont {Riccardi},\ and\ \citenamefont {Fontanesi}(1999)}]{400khz}%
  \BibitemOpen
  \bibfield  {author} {\bibinfo {author} {\bibfnamefont {G.}~\bibnamefont {Chiodini}}, \bibinfo {author} {\bibfnamefont {C.}~\bibnamefont {Riccardi}}, \ and\ \bibinfo {author} {\bibfnamefont {M.}~\bibnamefont {Fontanesi}},\ }\bibfield  {title} {\enquote {\bibinfo {title} {A 400 khz, fast-sweep langmuir probe for measuring plasma fluctuations},}\ }\href {\doibase 10.1063/1.1149828} {\bibfield  {journal} {\bibinfo  {journal} {Review of Scientific Instruments}\ }\textbf {\bibinfo {volume} {70}},\ \bibinfo {pages} {2681--2688} (\bibinfo {year} {1999})}\BibitemShut {NoStop}%
\bibitem [{\citenamefont {Horton}(1990)}]{fluctuations-energy-transport}%
  \BibitemOpen
  \bibfield  {author} {\bibinfo {author} {\bibfnamefont {W.}~\bibnamefont {Horton}},\ }\bibfield  {title} {\enquote {\bibinfo {title} {Fluctuations and anomalous transport in tokamaks},}\ }\href {\doibase 10.1063/1.859500} {\bibfield  {journal} {\bibinfo  {journal} {Physics of Fluids B: Plasma Physics}\ }\textbf {\bibinfo {volume} {2}},\ \bibinfo {pages} {2879--2903} (\bibinfo {year} {1990})}\BibitemShut {NoStop}%
\bibitem [{\citenamefont {Thomas}\ and\ \citenamefont {Cappelli}(1997)}]{thomas1997fluctuation}%
  \BibitemOpen
  \bibfield  {author} {\bibinfo {author} {\bibfnamefont {E.}~\bibnamefont {Thomas}}\ and\ \bibinfo {author} {\bibfnamefont {M.~A.}\ \bibnamefont {Cappelli}},\ }\bibfield  {title} {\enquote {\bibinfo {title} {Fluctuation-induced transport in the hall plasma accelerator},}\ }\href {\doibase 10.1063/1.872134} {\bibfield  {journal} {\bibinfo  {journal} {Physics of Plasmas}\ }\textbf {\bibinfo {volume} {4}},\ \bibinfo {pages} {447--457} (\bibinfo {year} {1997})}\BibitemShut {NoStop}%
\bibitem [{\citenamefont {Terrazas-Salinas}(2022)}]{tsf_test_plan}%
  \BibitemOpen
  \bibfield  {author} {\bibinfo {author} {\bibfnamefont {I.}~\bibnamefont {Terrazas-Salinas}},\ }\href {https://ntrs.nasa.gov/} {\enquote {\bibinfo {title} {Test planning guide for nasa ames research center arc jet complex and range complex},}\ }\bibinfo {type} {Tech. Rep.}\ \bibinfo {number} {A029-9701-XM3 Rev. J}\ (\bibinfo  {institution} {NASA Ames Research Center},\ \bibinfo {year} {2022})\ \bibinfo {note} {revision J}\BibitemShut {NoStop}%
\bibitem [{\citenamefont {Wang}\ and\ \citenamefont {Sun}(2015)}]{non-lte-arcjet}%
  \BibitemOpen
  \bibfield  {author} {\bibinfo {author} {\bibfnamefont {J.}~\bibnamefont {Wang}}\ and\ \bibinfo {author} {\bibfnamefont {Q.}~\bibnamefont {Sun}},\ }\bibfield  {title} {\enquote {\bibinfo {title} {Status and prospects on nonequilibrium modeling of high velocity plasma flow in an arcjet thruster},}\ }\href {\doibase 10.1007/s11090-015-9610-4} {\bibfield  {journal} {\bibinfo  {journal} {Plasma Chemistry and Plasma Processing}\ }\textbf {\bibinfo {volume} {35}},\ \bibinfo {pages} {925--962} (\bibinfo {year} {2015})}\BibitemShut {NoStop}%
\bibitem [{\citenamefont {Meurisse}\ \emph {et~al.}(2018)\citenamefont {Meurisse}, \citenamefont {Visser}, \citenamefont {Izquierdo}, \citenamefont {Haw},\ and\ \citenamefont {Mansour}}]{meurisse2018unsteady}%
  \BibitemOpen
  \bibfield  {author} {\bibinfo {author} {\bibfnamefont {J.~B.~E.}\ \bibnamefont {Meurisse}}, \bibinfo {author} {\bibfnamefont {S.~J.}\ \bibnamefont {Visser}}, \bibinfo {author} {\bibfnamefont {S.~F.}\ \bibnamefont {Izquierdo}}, \bibinfo {author} {\bibfnamefont {M.}~\bibnamefont {Haw}}, \ and\ \bibinfo {author} {\bibfnamefont {N.~N.}\ \bibnamefont {Mansour}},\ }\bibfield  {title} {\enquote {\bibinfo {title} {3d unsteady model of arc heater plasma flow using the arc heater simulator (arches)},}\ }in\ \href {https://ntrs.nasa.gov/api/citations/20190025359/downloads/20190025359.pdf} {\emph {\bibinfo {booktitle} {71st Annual Gaseous Electronics Conference \& 60th Annual Meeting of the APS Division of Plasma Physics}}}\ (\bibinfo {address} {Portland, Oregon},\ \bibinfo {year} {2018})\BibitemShut {NoStop}%
\bibitem [{\citenamefont {Guo}\ and\ \citenamefont {Wu}(2014)}]{Guo2014}%
  \BibitemOpen
  \bibfield  {author} {\bibinfo {author} {\bibfnamefont {H.}~\bibnamefont {Guo}}\ and\ \bibinfo {author} {\bibfnamefont {G.-Q.}\ \bibnamefont {Wu}},\ }\bibfield  {title} {\enquote {\bibinfo {title} {Three-dimensional non-equilibrium modeling on the characteristics of the dual-jet direct-current arc plasmas},}\ }\href@noop {} {\bibfield  {journal} {\bibinfo  {journal} {Plasma Chemistry and Plasma Processing}\ } (\bibinfo {year} {2014})},\ \bibinfo {note} {date of publication: October 10, 2014}\BibitemShut {NoStop}%
\bibitem [{\citenamefont {Baeva}\ \emph {et~al.}(2016)\citenamefont {Baeva}, \citenamefont {Benilov}, \citenamefont {Almeida},\ and\ \citenamefont {Uhrlandt}}]{Baeva2016}%
  \BibitemOpen
  \bibfield  {author} {\bibinfo {author} {\bibfnamefont {M.}~\bibnamefont {Baeva}}, \bibinfo {author} {\bibfnamefont {M.~S.}\ \bibnamefont {Benilov}}, \bibinfo {author} {\bibfnamefont {N.~A.}\ \bibnamefont {Almeida}}, \ and\ \bibinfo {author} {\bibfnamefont {D.}~\bibnamefont {Uhrlandt}},\ }\bibfield  {title} {\enquote {\bibinfo {title} {Novel non-equilibrium modelling of a dc electric arc in argon},}\ }\href {\doibase 10.1088/0022-3727/49/24/245205} {\bibfield  {journal} {\bibinfo  {journal} {Journal of Physics D: Applied Physics}\ }\textbf {\bibinfo {volume} {49}},\ \bibinfo {pages} {245205} (\bibinfo {year} {2016})}\BibitemShut {NoStop}%
\bibitem [{\citenamefont {Zube}\ and\ \citenamefont {Myers}(1993)}]{ZubeMyers1993}%
  \BibitemOpen
  \bibfield  {author} {\bibinfo {author} {\bibfnamefont {D.~M.}\ \bibnamefont {Zube}}\ and\ \bibinfo {author} {\bibfnamefont {R.~M.}\ \bibnamefont {Myers}},\ }\bibfield  {title} {\enquote {\bibinfo {title} {Thermal nonequilibrium in a low-power arcjet nozzle},}\ }\href {\doibase 10.2514/3.23657} {\bibfield  {journal} {\bibinfo  {journal} {Journal of Propulsion and Power}\ }\textbf {\bibinfo {volume} {9}} (\bibinfo {year} {1993}),\ 10.2514/3.23657}\BibitemShut {NoStop}%
\bibitem [{\citenamefont {Poulikakos}, \citenamefont {Zhao},\ and\ \citenamefont {Arcidiacono}(1996)}]{poulikakos1996electric}%
  \BibitemOpen
  \bibfield  {author} {\bibinfo {author} {\bibfnamefont {D.}~\bibnamefont {Poulikakos}}, \bibinfo {author} {\bibfnamefont {J.}~\bibnamefont {Zhao}}, \ and\ \bibinfo {author} {\bibfnamefont {M.}~\bibnamefont {Arcidiacono}},\ }\bibfield  {title} {\enquote {\bibinfo {title} {Electric arc fluctuations in dc plasma spray torch},}\ }\href {\doibase 10.1007/BF01570215} {\bibfield  {journal} {\bibinfo  {journal} {Plasma Chemistry and Plasma Processing}\ }\textbf {\bibinfo {volume} {16}},\ \bibinfo {pages} {1--23} (\bibinfo {year} {1996})}\BibitemShut {NoStop}%
\bibitem [{\citenamefont {Haw}\ \emph {et~al.}(2020)\citenamefont {Haw}, \citenamefont {Meurisse}, \citenamefont {Visser}, \citenamefont {Izquierdo}, \citenamefont {Schulz},\ and\ \citenamefont {Mansour}}]{ahf-bdot-probes}%
  \BibitemOpen
  \bibfield  {author} {\bibinfo {author} {\bibfnamefont {M.~A.}\ \bibnamefont {Haw}}, \bibinfo {author} {\bibfnamefont {J.~B.~E.}\ \bibnamefont {Meurisse}}, \bibinfo {author} {\bibfnamefont {S.~J.}\ \bibnamefont {Visser}}, \bibinfo {author} {\bibfnamefont {S.~F.}\ \bibnamefont {Izquierdo}}, \bibinfo {author} {\bibfnamefont {J.}~\bibnamefont {Schulz}}, \ and\ \bibinfo {author} {\bibfnamefont {N.~N.}\ \bibnamefont {Mansour}},\ }\bibfield  {title} {\enquote {\bibinfo {title} {Preliminary measurements of the motion of arcjet current channel using inductive magnetic probes},}\ }in\ \href {\doibase 10.2514/6.2020-0919} {\emph {\bibinfo {booktitle} {AIAA Scitech 2020 Forum}}}\ (\bibinfo {year} {2020})\BibitemShut {NoStop}%
\bibitem [{\citenamefont {Rodrigues}\ \emph {et~al.}(2024{\natexlab{a}})\citenamefont {Rodrigues}, \citenamefont {Macdonald}, \citenamefont {Haw}, \citenamefont {Martinez}, \citenamefont {Philippidis}, \citenamefont {Colom}, \citenamefont {Chung},\ and\ \citenamefont {Hartman}}]{Rodrigues2024}%
  \BibitemOpen
  \bibfield  {author} {\bibinfo {author} {\bibfnamefont {J.}~\bibnamefont {Rodrigues}}, \bibinfo {author} {\bibfnamefont {M.~E.}\ \bibnamefont {Macdonald}}, \bibinfo {author} {\bibfnamefont {M.~A.}\ \bibnamefont {Haw}}, \bibinfo {author} {\bibfnamefont {R.}~\bibnamefont {Martinez}}, \bibinfo {author} {\bibfnamefont {D.}~\bibnamefont {Philippidis}}, \bibinfo {author} {\bibfnamefont {S.}~\bibnamefont {Colom}}, \bibinfo {author} {\bibfnamefont {R.}~\bibnamefont {Chung}}, \ and\ \bibinfo {author} {\bibfnamefont {J.}~\bibnamefont {Hartman}},\ }\bibfield  {title} {\enquote {\bibinfo {title} {{Defining the Operational Envelope for Air Flows in the miniature Arc jet Research Chamber (mARC II)}},}\ }in\ \href {\doibase 10.2514/6.2024-3553} {\emph {\bibinfo {booktitle} {AIAA AVIATION FORUM AND ASCEND 2024}}}\ (\bibinfo  {publisher} {AIAA},\ \bibinfo {address} {Las Vegas, Nevada},\ \bibinfo {year} {2024})\BibitemShut {NoStop}%
\bibitem [{\citenamefont {Rodrigues}\ \emph {et~al.}(2025{\natexlab{a}})\citenamefont {Rodrigues}, \citenamefont {MacDonald}, \citenamefont {Haw}, \citenamefont {Martinez}, \citenamefont {Philippidis}, \citenamefont {Colom}, \citenamefont {Chung},\ and\ \citenamefont {Hartman}}]{marc-2025-expanding}%
  \BibitemOpen
  \bibfield  {author} {\bibinfo {author} {\bibfnamefont {J.}~\bibnamefont {Rodrigues}}, \bibinfo {author} {\bibfnamefont {M.~E.}\ \bibnamefont {MacDonald}}, \bibinfo {author} {\bibfnamefont {M.}~\bibnamefont {Haw}}, \bibinfo {author} {\bibfnamefont {R.}~\bibnamefont {Martinez}}, \bibinfo {author} {\bibfnamefont {D.}~\bibnamefont {Philippidis}}, \bibinfo {author} {\bibfnamefont {S.}~\bibnamefont {Colom}}, \bibinfo {author} {\bibfnamefont {R.~J.}\ \bibnamefont {Chung}}, \ and\ \bibinfo {author} {\bibfnamefont {J.}~\bibnamefont {Hartman}},\ }\bibfield  {title} {\enquote {\bibinfo {title} {Expanding the measurement capabilities of the marc ii arc-jet to map the operating envelope for high-enthalpy air flows},}\ }in\ \href {\doibase 10.2514/6.2025-2443} {\emph {\bibinfo {booktitle} {AIAA SciTech Forum}}}\ (\bibinfo {year} {2025})\BibitemShut {NoStop}%
\bibitem [{\citenamefont {Marshall}\ \emph {et~al.}(2025)\citenamefont {Marshall}, \citenamefont {Caldwell}, \citenamefont {Feldman}, \citenamefont {Chavez-Garcia}, \citenamefont {Hendrickson}, \citenamefont {Rodrigues}, \citenamefont {MacDonald}, \citenamefont {Turcotte}, \citenamefont {Alcantara}, \citenamefont {Swaiss},\ and\ \citenamefont {Bebak}}]{Marshall2025}%
  \BibitemOpen
  \bibfield  {author} {\bibinfo {author} {\bibfnamefont {P.}~\bibnamefont {Marshall}}, \bibinfo {author} {\bibfnamefont {A.}~\bibnamefont {Caldwell}}, \bibinfo {author} {\bibfnamefont {J.}~\bibnamefont {Feldman}}, \bibinfo {author} {\bibfnamefont {J.}~\bibnamefont {Chavez-Garcia}}, \bibinfo {author} {\bibfnamefont {K.}~\bibnamefont {Hendrickson}}, \bibinfo {author} {\bibfnamefont {J.}~\bibnamefont {Rodrigues}}, \bibinfo {author} {\bibfnamefont {M.}~\bibnamefont {MacDonald}}, \bibinfo {author} {\bibfnamefont {A.}~\bibnamefont {Turcotte}}, \bibinfo {author} {\bibfnamefont {T.}~\bibnamefont {Alcantara}}, \bibinfo {author} {\bibfnamefont {C.}~\bibnamefont {Swaiss}}, \ and\ \bibinfo {author} {\bibfnamefont {A.}~\bibnamefont {Bebak}},\ }\bibfield  {title} {\enquote {\bibinfo {title} {{Designing Insulative Reusable Thermal Protection Materials – from Shuttle Tile to Next Generation Systems}},}\ }in\ \href@noop {} {\emph {\bibinfo {booktitle} {Composites, Materials, {\&} Structures (CMS 2025) 48th Annual
  Conference}}}\ (\bibinfo {year} {2025})\BibitemShut {NoStop}%
\bibitem [{\citenamefont {Instruments}(2017)}]{78XX}%
  \BibitemOpen
  \bibfield  {author} {\bibinfo {author} {\bibfnamefont {T.}~\bibnamefont {Instruments}},\ }\href {https://www.ti.com/lit/ds/symlink/lm7800.pdf} {\emph {\bibinfo {title} {LM7800 Series 3-Terminal Positive Regulators}}} (\bibinfo {year} {2017}),\ \bibinfo {note} {datasheet, Texas Instruments}\BibitemShut {NoStop}%
\bibitem [{\citenamefont {Siliconix}(2021{\natexlab{a}})}]{vishay_irf620}%
  \BibitemOpen
  \bibfield  {author} {\bibinfo {author} {\bibfnamefont {V.}~\bibnamefont {Siliconix}},\ }\href {https://www.vishay.com/docs/91027/irf620.pdf} {\enquote {\bibinfo {title} {Irf620 power mosfet datasheet},}\ } (\bibinfo {year} {2021}{\natexlab{a}}),\ \bibinfo {note} {document Number: 91027}\BibitemShut {NoStop}%
\bibitem [{\citenamefont {Siliconix}(2021{\natexlab{b}})}]{vishay_irf9610}%
  \BibitemOpen
  \bibfield  {author} {\bibinfo {author} {\bibfnamefont {V.}~\bibnamefont {Siliconix}},\ }\href {https://www.vishay.com/docs/91080/91080.pdf} {\enquote {\bibinfo {title} {Irf9610 power mosfet datasheet},}\ } (\bibinfo {year} {2021}{\natexlab{b}}),\ \bibinfo {note} {document Number: 91080}\BibitemShut {NoStop}%
\bibitem [{\citenamefont {Rodrigues}\ \emph {et~al.}(2024{\natexlab{b}})\citenamefont {Rodrigues}, \citenamefont {MacDonald}, \citenamefont {Haw}, \citenamefont {Martinez}, \citenamefont {Philippidis}, \citenamefont {Colom}, \citenamefont {Chung},\ and\ \citenamefont {Hartman}}]{marc-2024-defining}%
  \BibitemOpen
  \bibfield  {author} {\bibinfo {author} {\bibfnamefont {J.}~\bibnamefont {Rodrigues}}, \bibinfo {author} {\bibfnamefont {M.~E.}\ \bibnamefont {MacDonald}}, \bibinfo {author} {\bibfnamefont {M.~A.}\ \bibnamefont {Haw}}, \bibinfo {author} {\bibfnamefont {R.}~\bibnamefont {Martinez}}, \bibinfo {author} {\bibfnamefont {D.}~\bibnamefont {Philippidis}}, \bibinfo {author} {\bibfnamefont {S.}~\bibnamefont {Colom}}, \bibinfo {author} {\bibfnamefont {R.}~\bibnamefont {Chung}}, \ and\ \bibinfo {author} {\bibfnamefont {J.}~\bibnamefont {Hartman}},\ }\bibfield  {title} {\enquote {\bibinfo {title} {Defining the operational envelope for air flows in the miniature arc-jet research chamber (marc ii)},}\ }in\ \href {\doibase 10.2514/6.2024-XXXX} {\emph {\bibinfo {booktitle} {AIAA Aviation Forum and Exposition}}}\ (\bibinfo {organization} {American Institute of Aeronautics and Astronautics},\ \bibinfo {address} {Las Vegas, NV},\ \bibinfo {year} {2024})\BibitemShut {NoStop}%
\bibitem [{\citenamefont {Bellan}(2006)}]{bellan2006fundamentals}%
  \BibitemOpen
  \bibfield  {author} {\bibinfo {author} {\bibfnamefont {P.~M.}\ \bibnamefont {Bellan}},\ }\href {https://www.cambridge.org/0521821169} {\emph {\bibinfo {title} {Fundamentals of Plasma Physics}}}\ (\bibinfo  {publisher} {Cambridge University Press},\ \bibinfo {address} {Cambridge, UK},\ \bibinfo {year} {2006})\BibitemShut {NoStop}%
\bibitem [{\citenamefont {Johnson}, \citenamefont {Weiser},\ and\ \citenamefont {Asmussen}(1997)}]{johnson1997electron}%
  \BibitemOpen
  \bibfield  {author} {\bibinfo {author} {\bibfnamefont {R.~E.}\ \bibnamefont {Johnson}}, \bibinfo {author} {\bibfnamefont {M.~S.}\ \bibnamefont {Weiser}}, \ and\ \bibinfo {author} {\bibfnamefont {J.}~\bibnamefont {Asmussen}},\ }\bibfield  {title} {\enquote {\bibinfo {title} {Electron densities and temperatures in a diamond-producing microwave plasma reactor},}\ }\href {\doibase 10.1063/1.364180} {\bibfield  {journal} {\bibinfo  {journal} {Journal of Applied Physics}\ }\textbf {\bibinfo {volume} {81}},\ \bibinfo {pages} {1093--1104} (\bibinfo {year} {1997})}\BibitemShut {NoStop}%
\bibitem [{\citenamefont {Lu\'{\i}s}\ and\ \citenamefont {MacDonald}(2021)}]{Diana2021}%
  \BibitemOpen
  \bibfield  {author} {\bibinfo {author} {\bibfnamefont {D.}~\bibnamefont {Lu\'{\i}s}}\ and\ \bibinfo {author} {\bibfnamefont {M.~E.}\ \bibnamefont {MacDonald}},\ }\bibfield  {title} {\enquote {\bibinfo {title} {Emission spectroscopy characterization of electrode species in the freestream flow at the nasa ames miniature arc jet ii facility},}\ }\href {\doibase 10.1016/j.jqsrt.2021.107752} {\bibfield  {journal} {\bibinfo  {journal} {Journal of Quantitative Spectroscopy and Radiative Transfer}\ }\textbf {\bibinfo {volume} {272}} (\bibinfo {year} {2021}),\ 10.1016/j.jqsrt.2021.107752}\BibitemShut {NoStop}%
\bibitem [{\citenamefont {Chen}(2020)}]{chen2020lecture}%
  \BibitemOpen
  \bibfield  {author} {\bibinfo {author} {\bibfnamefont {F.~F.}\ \bibnamefont {Chen}},\ }\href {https://www.seas.ucla.edu/~ffchen/Publs/Chen231R.pdf} {\enquote {\bibinfo {title} {Lecture notes on langmuir probe diagnostics},}\ }\bibinfo {type} {Tech. Rep.}\ (\bibinfo  {institution} {University of California, Los Angeles},\ \bibinfo {year} {2020})\ \bibinfo {note} {accessed: 2025-02-16}\BibitemShut {NoStop}%
\bibitem [{\citenamefont {Rodrigues}\ \emph {et~al.}(2025{\natexlab{b}})\citenamefont {Rodrigues}, \citenamefont {MacDonald}, \citenamefont {Haw}, \citenamefont {Martinez}, \citenamefont {Philippidis}, \citenamefont {Colom}, \citenamefont {Chung},\ and\ \citenamefont {Hartman}}]{Rodrigues2025a}%
  \BibitemOpen
  \bibfield  {author} {\bibinfo {author} {\bibfnamefont {J.}~\bibnamefont {Rodrigues}}, \bibinfo {author} {\bibfnamefont {M.~E.}\ \bibnamefont {MacDonald}}, \bibinfo {author} {\bibfnamefont {M.}~\bibnamefont {Haw}}, \bibinfo {author} {\bibfnamefont {R.}~\bibnamefont {Martinez}}, \bibinfo {author} {\bibfnamefont {D.}~\bibnamefont {Philippidis}}, \bibinfo {author} {\bibfnamefont {S.}~\bibnamefont {Colom}}, \bibinfo {author} {\bibfnamefont {R.~J.}\ \bibnamefont {Chung}}, \ and\ \bibinfo {author} {\bibfnamefont {J.}~\bibnamefont {Hartman}},\ }\bibfield  {title} {\enquote {\bibinfo {title} {{Expanding the Measurement Capabilities of the mARC II Arc-Jet to Map the Operating Envelope for High-Enthalpy Air Flows}},}\ }in\ \href {\doibase 10.2514/6.2025-2443} {\emph {\bibinfo {booktitle} {AIAA SCITECH 2025 Forum}}}\ (\bibinfo  {publisher} {American Institute of Aeronautics and Astronautics},\ \bibinfo {address} {Reston, Virginia},\ \bibinfo {year} {2025})\BibitemShut {NoStop}%
\bibitem [{\citenamefont {Palmer}\ \emph {et~al.}(2025)\citenamefont {Palmer}, \citenamefont {Prabhu}, \citenamefont {Driver},\ and\ \citenamefont {Rodrigues}}]{Palmer2025}%
  \BibitemOpen
  \bibfield  {author} {\bibinfo {author} {\bibfnamefont {G.}~\bibnamefont {Palmer}}, \bibinfo {author} {\bibfnamefont {D.}~\bibnamefont {Prabhu}}, \bibinfo {author} {\bibfnamefont {D.}~\bibnamefont {Driver}}, \ and\ \bibinfo {author} {\bibfnamefont {J.}~\bibnamefont {Rodrigues}},\ }\bibfield  {title} {\enquote {\bibinfo {title} {{CFD Simulations of Arcjet Experiments: A Survey of Recommended Practices and Lessons Learned}},}\ }in\ \href {\doibase https://arc.aiaa.org/doi/abs/10.2514/6.2025-3480} {\emph {\bibinfo {booktitle} {AIAA AVIATION 2025 FORUM}}}\ (\bibinfo  {publisher} {AIAA},\ \bibinfo {year} {2025})\BibitemShut {NoStop}%
\bibitem [{\citenamefont {Prevosto}, \citenamefont {Kelly},\ and\ \citenamefont {Mancinelli}(2008)}]{Prevosto_SSLP}%
  \BibitemOpen
  \bibfield  {author} {\bibinfo {author} {\bibfnamefont {L.}~\bibnamefont {Prevosto}}, \bibinfo {author} {\bibfnamefont {H.}~\bibnamefont {Kelly}}, \ and\ \bibinfo {author} {\bibfnamefont {B.}~\bibnamefont {Mancinelli}},\ }\bibfield  {title} {\enquote {\bibinfo {title} {On the use of sweeping langmuir probes in cutting arc plasmas—part i: Experimental results},}\ }\href {\doibase 10.1109/TPS.2007.914176} {\bibfield  {journal} {\bibinfo  {journal} {IEEE Transactions on Plasma Science}\ }\textbf {\bibinfo {volume} {36}},\ \bibinfo {pages} {263--270} (\bibinfo {year} {2008})}\BibitemShut {NoStop}%
\bibitem [{\citenamefont {Hairapetian}\ and\ \citenamefont {Stenzel}(1988)}]{HairapetianStenzel1988}%
  \BibitemOpen
  \bibfield  {author} {\bibinfo {author} {\bibfnamefont {G.}~\bibnamefont {Hairapetian}}\ and\ \bibinfo {author} {\bibfnamefont {R.~L.}\ \bibnamefont {Stenzel}},\ }\bibfield  {title} {\enquote {\bibinfo {title} {Expansion of a two-electron-population plasma into vacuum},}\ }\href {\doibase 10.1103/PhysRevLett.61.1607} {\bibfield  {journal} {\bibinfo  {journal} {Physical Review Letters}\ }\textbf {\bibinfo {volume} {61}},\ \bibinfo {pages} {1607--1610} (\bibinfo {year} {1988})}\BibitemShut {NoStop}%
\end{thebibliography}%

%%%%%%%%%%%%%%%%%%%%%%%%%%%%%%%%%%%%%%%%%%%%%%%%%%%%%%%%%%%%%%%%%%%%%%%%%%%%%%%%%%%%
\end{document}